\documentclass[12pt]{article}
\usepackage{geometry}
\usepackage{amssymb}
\usepackage{amsmath}
\usepackage{bbm}
\usepackage{MnSymbol}
\usepackage{hyperref}
\usepackage{braket}
\usepackage{cite}
\usepackage{xcolor}
\usepackage{subcaption,tikz}
\usetikzlibrary{arrows,decorations.markings,decorations.pathreplacing,decorations.shapes,decorations.pathmorphing}

\newcommand{\be}{\begin{equation}}
	\newcommand{\ee}{\end{equation}}
\newcommand{\bea}{\begin{eqnarray}}
	\newcommand{\eea}{\end{eqnarray}}

\newcommand{\al}{\alpha}
\renewcommand{\d}{\delta}
\newcommand{\e}{\epsilon}

\newcommand{\g}{\gamma}

\newcommand{\la}{\lambda}
\newcommand{\m}{\mu}

\newcommand{\mcL}{\mathcal{L}}
\newcommand{\mcC}{\mathcal{C}}

\newcommand{\mcT}{\mathcal{T}}

\newcommand{\C}{\mathbb{C}}

\newcommand{\hlf}{\frac{1}{2}}

\newcommand{\non}{\nonumber}

\newcommand{\Z}{\mathbb{Z}}

\newcommand{\id}{\mathbbm{1}}

\newcommand{\Hom}{\operatorname{Hom}}

\newcommand{\Tr}{\operatorname{Tr}}

\newcommand{\Rep}{\operatorname{Rep}}

\newcommand{\SPT}{\operatorname{SPT}}

\newcommand{\lp}{\left(}
\newcommand{\rp}{\right)}
\newcommand{\ls}{\left[}
\newcommand{\rs}{\right]}
\newcommand{\mcA}{\mathcal{A}}

\newcommand{\ov}[1]{{\overline{#1}}}

\newcommand{\op}{^\text{op}}

\def\sqr#1#2{{\vcenter{\vbox{\hrule height.#2pt
				\hbox{\vrule width.#2pt height#1pt \kern#1pt
					\vrule width.#2pt}\hrule height.#2pt}}}}

\tikzset{
	partial ellipse/.style args={#1:#2:#3}{
		insert path={+ (#1:#3) arc (#1:#2:#3)}
	}
}

\numberwithin{equation}{section}

\begin{document}
\begin{center}
	
{\large\bf The Fusion Categorical Diagonal}

\vspace*{0.2in}

Daniel Robbins$^1$, Thomas Vandermeulen$^2$

\vspace*{0.2in}

\begin{tabular}{cc}
	{\begin{tabular}{l}
			$^1$ Department of Physics\\
			University at Albany\\
			Albany, NY 12222 \end{tabular}}
	&
	{\begin{tabular}{l}
			$^2$ George P.~and Cynthia W.~Mitchell Institute\\
			for Fundamental Physics and Astronomy\\
			Texas A\&M University\\
			College Station, TX 77843 \end{tabular}}
\end{tabular}

\vspace*{0.2in}

{\tt dgrobbins@albany.edu, tvand@tamu.edu}

\end{center}
\pagenumbering{gobble}

We define a Frobenius algebra over fusion categories of the form Rep$(G)\boxtimes$Rep$(G)$ which generalizes the diagonal subgroup of $G\times G$.  This allows us to extend field theoretical constructions which depend on the existence of a diagonal subgroup to non-invertible symmetries.  We give explicit calculations for theories with Rep$(S_3)\boxtimes$Rep$(S_3)$ symmetry, applying the results to gauging topological quantum field theories which carry this non-invertible symmetry.  Along the way, we also discuss how Morita equivalence is implemented for algebras in symmetry categories.

\newpage
\tableofcontents

\newpage
	
\section{Introduction}
\pagenumbering{arabic}
\baselineskip=18pt
	
Diagonal symmetries have consistently proven useful in quantum field theory.  Perhaps the most prominent example lies in discrete torsion, which can be implemented by taking the direct product of a $G$-symmetric theory with a symmetry-protected topological (SPT) phase for that same $G$ symmetry.  When we gauge their diagonal $G$ symmetry, the topological twist in the SPT provides discrete torsion in the gauged theory.  The method of theta defects greatly generalizes this construction, now allowing the auxiliary theory to be an arbitrary $G$-symmetric topological quantum field theory (TQFT), not necessarily of the same dimension.  The assertion of \cite{UNI,Unifying} is that upon gauging the diagonal symmetry of this product theory, the auxiliary TQFT becomes a topological defect in the gauged theory, and one can achieve somewhat exotic results in this way (e.g.~this creates condensation defects in 3d theories).  In the case that the two theories are of the same dimension, one produces a `gauge defect' implementing a gauging of the gauged theory \cite{GaugeDefects}.\\

While the technique described above is powerful, it hinges on the fact that the direct product of two $G$-symmetric theories carries a natural $G$ symmetry -- the diagonal subgroup of $G\times G$.  Recent years, however, have taught us that not all symmetries are group-like, and one might rightfully wonder whether the method of theta defects extends to non-invertible symmetries.  For concreteness, we will specialize to two spacetime dimensions where general (not necessarily invertible) finite symmetries are known to be described by modular fusion categories.  The most naive possibility is that, given two theories with symmetry described by a fusion category $\mcC$, their product (which has symmetry\footnote{The $\boxtimes$ symbol indicates the Deligne tensor product defined on abelian categories (see, e.g.~\cite{EGNO}).  The details of the construction are not so important here, and it should be viewed as the minimal additional structure to ensure that the product is also a fusion category, for instance ensuring that the $\Hom$ sets are tensor products, not just Cartesian products.} $\mcC\boxtimes\mcC$) would always have a diagonal $\mcC$ subsymmetry.  Unfortunately, this is too hopeful, as we can easily produce counterexamples.

For instance, consider the fusion category Rep$(S_3)$, the simple objects of which are given by the three $S_3$ irreps: the trivial irrep 1 of dimension 1, the sign irrep $X$ of dimension 1 and the standard irrep $Y$ of dimension 2.  The fusion rules are given by the tensor product of irreps, which in this case is
\be
\label{rs3_fusion}
\begin{tabular}{c | c c c}
	$\otimes$ & $1$ & $X$ & $Y$ \\\cline{1-4}
	$1$ & $1$ & $X$ & $Y$ \\
	$X$ & $X$ & $1$ & $Y$ \\
	$Y$ & $Y$ & $Y$ & $1+X+Y$ \\
\end{tabular}.
\ee
The direct product of two Rep$(S_3)$-symmetric theories would have symmetry $\Rep(S_3)\boxtimes\Rep(S_3)$, whose fusion rules are two independent copies of (\ref{rs3_fusion}).  Does such a theory have a diagonal Rep$(S_3)$ symmetry?  The diagonal elements are $(1,1)$, $(X,X)$ and $(Y,Y)$, but we can immediately see that these do not span a Rep$(S_3)$ subcategory by calculating
\begin{multline}
(Y,Y)\otimes(Y,Y) = (1,1) + (1,X) + (1,Y) 
\\+ (X,1) + (X,X) + (X,Y) + (Y,1) + (Y,X) + (Y,Y)
\end{multline}
which contains non-diagonal elements.  This will be a general problem -- diagonal elements in fusion categories will not always close under fusion.  Another way to see the general failure of the existence of a diagonal subcategory is that a simple object $(\mcL,\mcL)$ has weight $|\mcL|^2$ rather than $|\mcL|$, so the diagonal elements of $\mcC\boxtimes\mcC$ will not have the correct weights to form a subcategory isomorphic to $\mcC$.\\

A workaround for this problem was suggested in \cite[appendix B]{Lin_2023} and \cite[appendix A.3]{GaugeDefects}, which we will now review.  Let $A$ and $B$ be two 2d theories with Rep$(G)$ symmetry, where $G$ is a finite group (not necessarily abelian).  There should exist a special symmetric Frobenius algebra $\mcA$ in $\Rep(G)\boxtimes\Rep(G)$ such that
\be
\label{fc_diag}
\frac{A\otimes B}{\mcA} = \frac{A/\Rep(G)\otimes B/\Rep(G)}{G}.
\ee
Here on the right-hand side $A$ and $B$ each appear gauged by the regular representation of $G$.  This produces two theories with quantum symmetry $G$, so we can sensibly gauge the diagonal $G$ subgroup of their product.  The claim is that (\ref{fc_diag}), which makes sense whether $G$ is abelian or non-abelian, provides an appropriate generalization of the diagonal subgroup to Rep$(G)$ symmetries.  This algebra $\mcA$ in fact appears in \cite{EGNO} in a slightly broader context where it is referred to as the \textit{canonical algebra} of a tensor category.  The aim of this paper will be to give, in the specialized case of Rep$(G)$ fusion categories, concrete partition function computations (along the lines of and using the technology developed in \cite{Fuchs_2002,Notes1}) for gauging this symmetry.  Because of its duality (\ref{fc_diag}) to the diagonal subgroup, we will refer to the canonical algebra $\mcA$ in this context as the \textit{diagonal algebra}.

In section~\ref{sec:AbelianCase} we introduce the construction in the case where $G$ is abelian, and so $\Rep(G)$ is again group-like (and isomorphic to $G$).  In section~\ref{sec:diag} we proceed to the general case where $G$ is non-abelian and hence $\Rep(G)$ includes non-invertible symmetries.  We work through the case $G=S_3$ in some detail.  Section~\ref{sec:tqft} applies this procedure to the case where one of the theories is a TQFT, and demonstrates how all the possible gaugings of a theory with $\Rep(G)$ symmetry can be obtained in this way.  We conclude in section~\ref{sec:discussion} with some conclusions and discussion of next steps.  Finally, some extra details are provided in the appendices.  Appendix~\ref{app:folding} gives an alternative perspective on our construction of the diagonal algebra using the ``folding trick''.  Appendix~\ref{app:RepS3Details} provides, for completeness, some details of $\Rep(S_3)$ gauging.  Appendix~\ref{app:PTRelations} derives a relation among $\Rep(S_3)$ partial traces by nucleating a bubble and repeatedly using fusion moves.  And finally, appendix~\ref{app:MoritaEquivalentAlgebra} explains how this bubble nucleation is an example of Morita equivalence between algebras, and reviews how this construction proceeds more generally.

\section{Specialization to Abelian Case}
\label{sec:AbelianCase}

To begin with, we should check that (\ref{fc_diag}) makes sense when $G$ is an abelian group.  In this case the Rep$(G)$ symmetry is group-like and non-canonically isomorphic to $G$, so we write it as $\hat{G}$ to distinguish the two.  Then the diagonal algebra $\mcA$ should reduce to the diagonal $\hat{G}\in\hat{G}\times\hat{G}$ subgroup, and (\ref{fc_diag}) should take the form
\be
\label{abelian_diagonal}
\frac{A\otimes B}{\hat{G}} = \frac{A/\hat{G}\otimes B/\hat{G}}{G}.
\ee
We will argue for (\ref{abelian_diagonal}) by calculations at the level of genus one partition functions, which will be a straightforward exercise in dual (quantum) symmetries.\\

As a quick review, let us detail how gauging a quantum symmetry works in a $\Z_2$ orbifold.  Let $A$ be a theory with non-anomalous $\Z_2$ symmetry, and let $A_{g_1,g_2}$ be the twisted genus one partition functions of the orbifold theory (i.e.~the vacuum correlation function for the theory on a torus with $g_1$- and $g_2$-twisted boundary conditions along the homotopy cycles) -- we will refer to these objects as \textit{partial traces}.  The partition function obtained when gauging $A$'s $\Z_2$ symmetry is (writing $\Z_2$ additively)
\be
Z(A/\Z_2)=\frac{1}{2}[A_{0,0} + A_{0,1} + A_{1,0} + A_{1,1}].
\ee
The resulting gauged theory has a natural $\hat{\Z}_2$ quantum/dual symmetry which acts by a minus sign on the $\Z_2$-twisted states of the original theory (which are now genuine local operators after gauging).  Thus we can write the orbifold partition function with an insertion of the non-trivial element of $\hat{\Z}_2$ as
\be
A'_{0,1} = \frac{1}{2}[A_{0,0} + A_{0,1} - A_{1,0} - A_{1,1}].
\ee
By modular transformations we can easily fill out the other two partial traces in the orbit:
\begin{align}
A'_{1,0} &= \frac{1}{2}[A_{0,0} - A_{0,1} + A_{1,0} - A_{1,1}]\\
A'_{1,1} &= \frac{1}{2}[A_{0,0} - A_{0,1} - A_{1,0} + A_{1,1}].
\end{align}
These relations, along with $A'_{0,0} = Z(A/\Z_2)$, give us
\be
Z([A/\Z_2]/\hat{\Z}_2)=\frac{1}{2}[A'_{0,0} + A'_{0,1} + A'_{1,0} + A'_{1,1}] = A_{0,0} = Z(A)
\ee
and we see that gauging the dual symmetry has returned the original theory, as claimed. \\

We can easily repeat this calculation for an arbitrary abelian group $G$.  Letting $A$ now have symmetry $G$, we know that (here we will switch to multiplicative notation)
\be
Z(A/G) = \frac{1}{|G|}\sum_{g_1,g_2\in G}A_{g_1,g_2}.
\ee
The general form of the partial traces when gauging the quantum symmetry will be
\be
\label{qs_sectors}
A'_{\hat{g}_1,\hat{g}_2} = \frac{1}{|G|}\sum_{g_3,g_4\in G}\chi_{\hat{g}_2}(g_3)\chi_{\hat{g}_1}(g^{-1}_4)A_{g_3,g_4}
\ee
such that the full $\hat{G}$ orbifold is
\begin{multline}
Z([A/G]/\hat{G}) = \frac{1}{|G|^2}\sum_{\substack{\hat{g}_1,\hat{g}_2\in\hat{G} \\ g_3,g_4\in G}}\chi_{\hat{g}_2}(g_3)\chi_{\hat{g}_1}(g^{-1}_4)A_{g_3,g_4}\\
=\frac{1}{|G|^2}\sum_{g_3,g_4\in G}A_{g_3,g_4}\left(\sum_{\hat{g}_1\in\hat{G}}\chi_{\hat{g}_1}(g_4^{-1})\chi_{\hat{g}_1}(1)\right)\left(\sum_{\hat{g}_2\in\hat{G}}\chi_{\hat{g}_2}(g_3)\chi_{\hat{g}_2}(1)\right)\\
=\frac{1}{|G|^2}\sum_{g_3,g_4\in G}A_{g_3,g_4}\cdot|G|^2\delta_{g_3,1}\delta_{g_4,1} = Z_{1,1} = Z(A)
\end{multline}
where the sums over $\hat{g}_1$ and $\hat{g}_2$ are evaluated by character orthogonality (and the triviality of characters evaluated on the identity element allowed us to insert the extra factors).  Again, the $\hat{G}$ orbifold of the $G$ orbifold returns the original theory.\\

Now we move to the actual case of interest: two $G$-symmetric theories $A$ and $B$.  In order to test whether (\ref{abelian_diagonal}) holds, we would like to calculate the partition function\footnote{It's arbitrary which of the symmetries we label $G$ vs.~$\hat{G}$, so in keeping with the previous calculation we will let the first orbifold be by $G$ and the second be by $\hat{G}$.}
\be
Z\left(\frac{A/G\otimes B/G}{\hat{G}}\right).
\ee
Letting $A_{g_1,g_2}$ and $B_{g_1,g_2}$ be the partial traces of the $G$ orbifolds of these two theories, the sectors of the $\hat{G}\times\hat{G}$ orbifold of their product takes the form
\be
(A'\otimes B')_{\hat{g}_1,\hat{g}_2,\hat{g}_3,\hat{g}_4} = \frac{1}{|G|^2}\left[\sum_{g_5,g_6}\chi_{\hat{g}_2}(g_5)\chi_{\hat{g}_1}(g_6^{-1})A_{g_5,g_6}\right]\left[\sum_{g_7,g_8}\chi_{\hat{g}_4}(g_7)\chi_{\hat{g}_3}(g_8^{-1})B_{g_7,g_8}\right].
\ee
We can get the partition function for the diagonal $\hat{G}$ gauging by setting $\hat{g}_1=\hat{g}_3$, $\hat{g}_2=\hat{g}_4$, dividing by $|G|$ and summing:
\begin{multline}
Z\left(\frac{A/G\otimes B/G}{\hat{G}}\right) = \frac{1}{|G|^3}\sum_{\hat{g}_1,\hat{g}_2}\left[\sum_{g_3,g_4}\chi_{\hat{g}_2}(g_3)\chi_{\hat{g}_1}(g_4^{-1})A_{g_3,g_4}\right]\left[\sum_{g_5,g_6}\chi_{\hat{g}_2}(g_5)\chi_{\hat{g}_1}(g_6^{-1})B_{g_5,g_6}\right]\\
=\frac{1}{|G|^3}\left[\sum_{g_3,g_4}A_{g_3,g_4}\right]\left[\sum_{g_5,g_6}B_{g_5,g_6}\right]\left[\sum_{\hat{g}_1}\chi_{\hat{g}_1}(g_4^{-1})\chi_{\hat{g}_1}(g_6^{-1})\right]\left[\sum_{\hat{g}_2}\chi_{\hat{g}_2}(g_3)\chi_{\hat{g}_2}(g_5)\right]\\
= \frac{1}{|G|^3}\left[\sum_{g_3,g_4}A_{g_3,g_4}\right]\left[\sum_{g_5,g_6}B_{g_5,g_6}\right]|G|^2\delta_{g_4^{-1},g_6}\delta_{g_3^{-1},g_5} = \frac{1}{|G|}\sum_{g_3,g_4}A_{g_3,g_4}B_{g_3^{-1},g_4^{-1}} \\
= \frac{1}{|G|}\sum_{g_3,g_4}A_{g_3,g_4}B_{g_3,g_4} = Z\left(\frac{A\otimes B}{G}\right)
\end{multline}
which confirms (\ref{abelian_diagonal}), at least at the level of torus partition functions.  Note that we used the fact that $B_{g_3^{-1},g_4^{-1}}=B_{g_3,g_4}$, which holds because we can make a coordinate change $z\mapsto -z$ in the 2D spacetime.

\section{The Diagonal Algebra}
\label{sec:diag}

In this section we will assume that $G$ is nonabelian, forcing us to work with Frobenius algebras over the fusion categories Rep$(G)$ and $\Rep(G)\boxtimes\Rep(G)$.  First note that for finite groups $G$ and $H$,  $\Rep(G)\boxtimes\Rep(H)$ is (non-canonically) isomorphic to $\Rep(G\times H)$.  The irreps of $G\times H$ are in one-to-one correspondence with tensor products of irreps of $G$ and $H$.

In \cite{Notes1} it was shown by explicit construction that there exists a Frobenius algebra $\mathcal{A}$ of $\Rep(G)$ corresponding to a subgroup $H\subset G$ (with a choice of discrete torsion), such that $(T/G)/\mathcal{A}\cong T/H$.  The algebra object was simply the representation whose basis vectors $v_{gH}$ were labeled by cosets $gH\in G/H$, with $G$ action given by $\rho(k)\cdot v_{gH}=v_{kgH}$.  The algebra multiplication was just given by $\m(v_{gH}\otimes v_{kH})=\d_{gH,kH}\,v_{gH}$.

So to find the ``diagonal'' algebra of $\Rep(G)\boxtimes\Rep(G)$, all we need to do is apply this coset construction to the diagonal subgroup $G_{diag}\subset G\times G$.  The cosets in $(G\times G)/G_{diag}$ are easy to describe; we can choose representatives $(g,1)G_{diag}$, so our representation $(V,\rho)$ will have basis vectors $v_g:=v_{(g,1)G_{diag}}$, with the $G\times G$ action $\rho(g,h)\cdot v_k=v_{(gk,h)G_{diag}}=v_{(gkh^{-1},1)G_{diag}}=v_{gkh^{-1}}$.  To see how this decomposes into irreps of $G\times G$, we consider the character.  Since $\rho$ acts as a permutation on the basis vectors $v_g$, this is just a counting problem.  We find
\be
\chi(g,h)=\Tr(\rho(g,h))=\left\{\begin{matrix} 0, & \mathrm{if\ }h\notin\mathcal{C}_g, \\ \frac{|G|}{|\mathcal{C}_g|}, & \mathrm{if\ }h\in\mathcal{C}_g, \end{matrix}\right.
\ee
where $\mathcal{C}_g=\{kgk^{-1}\mid k\in G\}$ is the conjugacy class of $g$.  To then determine the decomposition of $(V,\rho)$ into irreps $(U_i\otimes U_j,\rho_i\otimes\rho_j)$, where $i,j$ label irreps of $G$, we take inner products.
\begin{align}
	(\chi,\chi_i\otimes\chi_j)=\ & \frac{1}{|G|^2}\sum_{g,h\in G}\chi(g,h)\overline{\chi_i(g)}\overline{\chi_j(h)}\non\\
	=\ & \frac{1}{|G|^2}\sum_{\mathcal{C}_g,\mathcal{C}_h\subset G}|\mathcal{C}_g||\mathcal{C}_h|\chi(g,h)\overline{\chi_i(g)}\overline{\chi_j(h)}\non\\
	=\ & \frac{1}{|G|^2}\sum_{\mathcal{C}_g\subset G}|\mathcal{C}_g|^2\frac{|G|}{|\mathcal{C}_g|}\overline{\chi_i(g)}\overline{\chi_j(g)}\non\\
	=\ & \frac{1}{|G|}\sum_{\mathcal{C}_g\subset G}|\mathcal{C}_g|\overline{\chi_i(g)}\overline{\chi_j(g)}=\delta_{i,\bar{\jmath}}.
\end{align}
That is, the representation $(U_i\otimes U_j,\rho_i\otimes\rho_j)$ only appears if $(U_j,\rho_j)$ is the conjugate representation to $(U_i,\rho_i)$, and in that case it occurs with multiplicity one.  So, changing notation a little, we can write
\be
\label{algob_rho}
\rho\cong\bigoplus_{\mathrm{irreps\ }i\mathrm{\ of\ }G}\rho_i\otimes\overline{\rho_i}.
\ee
The multiplication is given by
\be
\m(v_g\otimes v_h)=\d_{g,h}\,v_g.
\ee
And the unit, co-unit, and co-multiplication are
\begin{align}
	u(1)=\ & \sum_{g\in G}v_g,\\
	c(v_g)=\ & 1,\\
	\Delta(v_g)=\ & v_g\otimes v_g.
\end{align}
In fact, this is isomorphic to the Frobenius algebra corresponding to the regular representation of $G$, but now we are viewing it as an algebra inside the larger fusion category $\Rep(G\times G)\cong\Rep(G)\boxtimes\Rep(G)$.  From (\ref{algob_rho}) we can see that its algebra object is given by
\be
\sum_{\mcL\in\Rep(G)}(\mcL,\ov{\mcL})
\ee
where the sum runs over simple objects (i.e.~irreps).  This agrees with the description of the canonical algebra given in \cite[Definition 7.9.12]{EGNO}.\footnote{Note that the fact that the representation on the rhs of the tensor product in (\ref{algob_rho}) appears conjugated reflects the fact that this is an algebra on $\mcC\boxtimes\mcC\op$ for a fusion category $\mcC$, where the opposite category $\mcC\op$ is obtained by reversing the directions of the morphisms in $\mcC$.  This is reflected in the group-like case in e.g.~(\ref{qs_sectors}) where $g_3$ appears alongside $g_4^{-1}$.  Since our primary example Rep$(S_3)$ involves only self-dual objects, we will gloss over this fact going forward.}

\subsection{Example: $\Rep(S_3)\times\Rep(S_3)$}
\label{sec:rs3rs3}

Applying this construction to the $\Rep(S_3)$ case, our algebra object is
\be
\mathcal{A}=(1,1)+(X,X)+(Y,Y).
\ee
The quantum dimension of $\mathcal{A}$ is $1^2+1^2+2^2=6$.

\subsubsection{$S_3$ Gauging}
\label{sec:s3_gauge}

For later use, let's review how the partition functions look when gauging $S_3\times S_3$ or its diagonal $S_3$ subgroup.  Suppose we start with $A\otimes B$, where $A$ and $B$ are theories that each have non-anomalous $S_3$ symmetry.  Gauging the $S_3\times S_3$ symmetry group gives us $(A/S_3)\otimes(B/S_3)$, with partition function
\begin{align}
	Z_{S_3\times S_3}=\ & \frac{1}{36}\sum_{\substack{g,h,g',h'\in S_3 \\ gh=hg, g'h'=h'g'}}Z^{(A)}_{g,h}Z^{(B)}_{g',h'}\non\\
	=\ & \lp\frac{1}{6}\sum_{\substack{g,h\in S_3 \\ gh=hg}}Z^{(A)}_{g,h}\rp\lp\frac{1}{6}\sum_{\substack{g',h'\in S_3 \\ g'h'=h'g'}}Z^{(B)}_{g',h'}\rp\non\\
	=\ & \frac{1}{36}\lp Z^{(A)}_{1,1}+3Z^{(A)}_{1,a}+2Z^{(A)}_{1,b}+3Z^{(A)}_{a,1}+3Z^{(A)}_{a,a}+2Z^{(A)}_{b,1}+2Z^{(A)}_{b,b}+2Z^{(A)}_{b,b^2}\rp\non\\
	& \quad\times\lp Z^{(B)}_{1,1}+3Z^{(B)}_{1,a}+2Z^{(B)}_{1,b}+3Z^{(B)}_{a,1}+3Z^{(B)}_{a,a}+2Z^{(B)}_{b,1}+2Z^{(B)}_{b,b}+2Z^{(B)}_{b,b^2}\rp
\end{align}
On the other hand, if we had only gauged the diagonal $S_3$, we would get partition function
\begin{align}
\label{eq:S3DiagPF}
	Z_{(S_3)_\text{diag.}}=\ & \frac{1}{6}\sum_{\substack{g,h\in S_3 \\ gh=hg}}Z^{(A)}_{g,h}Z^{(B)}_{g,h}\non\\
	=\ & \frac{1}{6}\lp Z^{(A)}_{1,1}Z^{(B)}_{1,1}+3Z^{(A)}_{1,a}Z^{(B)}_{1,a}+2Z^{(A)}_{1,b}Z^{(B)}_{1,b}+3Z^{(A)}_{a,1}Z^{(B)}_{a,1}+3Z^{(A)}_{a,a}Z^{(B)}_{a,a}\right.\non\\
	& \qquad\left. +2Z^{(A)}_{b,1}Z^{(B)}_{b,1}+2Z^{(A)}_{b,b}Z^{(B)}_{b,b}+2Z^{(A)}_{b,b^2}Z^{(B)}_{b,b^2}\rp.
\end{align}

\subsubsection{$\Rep(S_3)$ Gauging}
\label{subsubsec:RepSGauging}

Let's briefly recall how we approach a theory with $\Rep(S_3)$ symmetry, following~\cite{Notes1}.  There are three irreps in $\Rep(S_3)$, labeled $1$, $X$, and $Y$, and they can be taken to act via
\be
\rho_1(a)=1,\qquad\rho_1(b)=1,
\ee
\be
\rho_X(a)=-1,\qquad\rho_X(b)=1,
\ee
\be
\rho_Y(a)=\lp\begin{matrix} -1 & 0 \\ 0 & 1 \end{matrix}\rp,\qquad\rho_Y(b)=\lp\begin{matrix} -\hlf & -\frac{\sqrt{3}}{2} \\ \frac{\sqrt{3}}{2} & -\hlf \end{matrix}\rp.
\ee
We will denote the basis vectors corresponding to these representations as $\{e\}$ for $1$, $\{e_X\}$ for $X$, and $\{e_1,e_2\}$ for $Y$.

Fusion corresponds to taking tensor product representations and obeys (\ref{rs3_fusion}).  To proceed we should choose a particular basis $\la_{R_1,R_2}^{R_3}\in\Hom(R_1\otimes R_2,R_3)$ for the intertwiners implementing these fusions.  When the trivial representation is involved there is a canonical choice, and we write\footnote{To save space we leave the tensor product symbol implicit, but an expression like $ee_X$ should be understood as the vector $e\otimes e_X$ inside the representation $1\otimes X$.}
\be
\la_{1,1}^1(ee)=e,\quad\la_{1,X}^X(ee_X)=\la_{X,1}^X(e_Xe)=e_X,\quad\la_{1,Y}^Y(ee_a)=\la_{Y,1}^Y(e_ae)=e_a,\ a=1,2.
\ee
For the other fusions, the choices are not canonical, but the ambiguity can be parameterized by $\C^\times$ valued numbers $\beta_1,\cdots,\beta_6$, as in~\cite[section 3.1]{Notes1}.\footnote{To match up with the choice of gauge used in e.g.~\cite{Chang_2019}, one would set $\beta_2 = \beta_4 = \beta_5 = 1$ and $\beta_1 = \beta_3 = \beta_6 =-1$.}  Also following the conventions in that paper, our choice for the fusion intertwiner basis can be used to determine a basis for the co-fusion intertwiners, $\d_{R_1}^{R_2,R_3}\in\Hom(R_1,R_2\otimes R_3)$.  For completeness we list these in Appendix~\ref{app:RepS3Details}.  These can be used to compute the components of the associator, but we won't explicitly need those details.

Now to gauge the full $\Rep(S_3)$, we construct a Frobenius algebra corresponding to the regular representation $1+X+2Y$.  Taking a basis of vectors $v_g$, $g\in S_3$, and the obvious left-action $\rho_{reg}(g)\cdot v_h=v_{gh}$, we can decompose into irreps,
\begin{align}
e=\ & v_1+v_b+v_{b^2}+v_a+v_{ab}+v_{ab^2},\\
e_X=\ & \frac{c_X}{\sqrt{6}}\lp v_1+v_b+v_{b^2}-v_a-v_{ab}-v_{ab^2}\rp,\\
e_{Y_11}=\ & \frac{c_1}{2}\lp v_b-v_{b^2}-v_{ab}+v_{ab^2}\rp,\\
e_{Y_12}=\ & \frac{c_1}{2\sqrt{3}}\lp -2v_1+v_b+v_{b^2}-2v_a+v_{ab}+v_{ab^2}\rp,\\
e_{Y_21}=\ & \frac{c_2}{2\sqrt{3}}\lp -2v_1+v_b+v_{b^2}+2v_a-v_{ab}-v_{ab^2}\rp,\\
e_{Y_22}=\ & \frac{c_2}{2}\lp -v_b+v_{b^2}-v_{ab}+v_{ab^2}\rp.
\end{align}
The multiplication is $\m(v_gv_h)=\d_{g,h}v_g$ (note that it is symmetric) and the co-multiplication is $\Delta(v_g)=v_gv_g$.  In Appendix~\ref{app:RepS3Details} we write these out explicitly in the basis of $e$'s to read off components $\mu_{A,B}^{C}$ and $\Delta_{C}^{B,A}$, where $A,B,C$ are the simple representations appearing in the decomposition of the algebra object $1+X+2Y$ (i.e.~one trivial representation, one $X$ representation, and two different $Y$ representations).  Finally, one gets the orbifold partition function from the formula $Z=\sum_{A,B,C}\m_{A,B}^C\Delta_C^{B,A}Z_{A,B}^C$.  In the current case this results in (combining $Y_1$ and $Y_2$ terms together)
\begin{align}
\label{rs3pf}
Z_{1+X+2Y}=\ & \frac{1}{6}\ls Z_{1,1}^1+Z_{1,X}^X+2Z_{1,Y}^Y+Z_{X,1}^X+Z_{X,X}^1-\frac{2\beta_1}{\beta_2\beta_3}Z_{X,Y}^Y\right.\non\\
& \qquad\left. +2Z_{Y,1}^Y-\frac{2\beta_4}{\beta_3\beta_6}Z_{Y,X}^Y+2Z_{Y,Y}^1+\frac{2\beta_4}{\beta_2\beta_6}Z_{Y,Y}^X+\frac{2\beta_4}{\beta_5^2}Z_{Y,Y}^Y\rs.
\end{align}

\subsubsection{$\Rep(S_3\times S_3)$ Gauging}

Now let us turn to the product, $\Rep(S_3\times S_3)$.  In this case there are nine irreps, but we will only worry about the diagonal ones, which we will denote as $11$, $XX$, and $YY$.  The corresponding representations are obtained from $\rho_{11}(g,h)=1$,
\be
\rho_{XX}(1,a)=\rho_{XX}(a,1)=-1,\qquad\rho_{XX}(1,b)=\rho_{XX}(b,1)=1,
\ee
and
\be
\rho_{YY}(1,a)=\lp\begin{matrix} -1 & 0 & 0 & 0 \\ 0 & 1 & 0 & 0 \\ 0 & 0 & -1 & 0 \\ 0 & 0 & 0 & 1 \end{matrix}\rp,\qquad\rho_{YY}(a,1)=\lp\begin{matrix} -1 & 0 & 0 & 0 \\ 0 & -1 & 0 & 0 \\ 0 & 0 & 1 & 0 \\ 0 & 0 & 0 & 1 \end{matrix}\rp,
\ee
\be
\rho_{YY}(1,b)=\lp\begin{matrix} -\hlf & -\frac{\sqrt{3}}{2} & 0 & 0 \\ \frac{\sqrt{3}}{2} & -\hlf & 0 & 0 \\ 0 & 0 & -\hlf & -\frac{\sqrt{3}}{2} \\ 0 & 0 & \frac{\sqrt{3}}{2} & -\hlf \end{matrix}\rp,\qquad\rho_{YY}(b,1)=\lp\begin{matrix} -\hlf & 0 & -\frac{\sqrt{3}}{2} & 0 \\ 0 & -\hlf & 0 & -\frac{\sqrt{3}}{2} \\ \frac{\sqrt{3}}{2} & 0 & -\hlf & 0 \\ 0 & \frac{\sqrt{3}}{2} & 0 & -\hlf \end{matrix}\rp.
\ee
In this case we'll denote the bases as $\{e\}$ for $11$, $\{e_{XX}\}$ for $XX$, and $\{e_{11},e_{12},e_{21},e_{22}\}$ for $YY$.  For future reference, we note that these are simply the tensor products of two copies of $\Rep(S_3)$ irreps, so for example
\be
\rho_{YY}(e_{ij})=\rho_Y(e_i)\otimes\rho_Y(e_j).
\ee
We again need to pick bases for the fusion intertwiners.  Restricting to only fusions involving these diagonal representations, we have canonical fusions for $11$, and for the others (here we take $YY$ indices mod 2)
\begin{align}
\la_{XX,XX}^11(e_{XX}e_{XX})=\ & \beta_1e,\\
\la_{XX,YY}^{YY}(e_{XX}e_{ij})=\ & (-1)^{i+j}\beta_2e_{i+1,j+1},\\
\la_{YY,XX}^{YY}(e_{11}e_{XX})=\ & (-1)^{i+j}\beta_3e_{i+1,j+1},\\
\la_{YY,YY}^{11}(e_{ij}e_{k\ell})=\ & \d_{ik}\d_{j\ell}\beta_4e,\\
\la_{YY,YY}^{XX}(e_{ij}e_{k\ell})=\ & (-1)^{i+j}\d_{i,k+1}\d_{j,\ell+1}\beta_6e_{XX},\\
\la_{YY,YY}^{YY}(e_{ij}e_{k\ell})=\ & (-1)^{(i+1)(k+1)+(j+1)(\ell+1)}\beta_5e_{i+k,j+\ell}.
\end{align}
In fact, all of these cases could be obtained from a general expression (in hopefully obvious notation)
\be
\la_{AB,CD}^{EF}(v_{AB}v_{CD})=\la_{A,C}^E(v_Av_C)\otimes\la_{B,D}^F(v_Bv_D).
\ee
From this perspective, the $\beta_i$ above are the products of $\Rep(S_3)$ $\beta_i$, i.e.~
\be
\beta_i=\beta_i'\beta_i''.
\ee
We can construct a basis of co-fusion intertwiners by following our established procedure, and it ends up working the same way, i.e.~
\be
\d_{EF}^{CD,AB}(v_{EF})=\d_E^{C,A}(v_E)\otimes\d_F^{D,B}(v_F).
\ee
In particular for our case
\begin{align}
\d_{11}^{XX,XX}(e)=\ & \beta_1^{-1}e_{XX}e_{XX},\\
\d_{11}^{YY,YY}(e)=\ & \beta_4^{-1}\lp e_{11}e_{11}+e_{12}e_{12}+e_{21}e_{21}+e_{22}e_{22}\rp,\\
\d_{XX}^{YY,YY}(e_{XX})=\ & \beta_2\beta_4^{-1}\lp e_{11}e_{22}-e_{12}e_{21}-e_{21}e_{12}+e_{22}e_{11}\rp,\\
\d_{YY}^{XX,YY}(e_{11})=\ & \beta_6\beta_4^{-1}e_{XX}e_{22},\\
\d_{YY}^{XX,YY}(e_{12})=\ & -\beta_6\beta_4^{-1}e_{XX}e_{21},\\
\d_{YY}^{XX,YY}(e_{21})=\ & -\beta_6\beta_4^{-1}e_{XX}e_{12},\\
\d_{YY}^{XX,YY}(e_{22})=\ & \beta_6\beta_4^{-1}e_{XX}e_{11},\\
\d_{YY}^{YY,XX}(e_{11})=\ & \beta_3\beta_1^{-1}e_{22}e_{XX},\\
\d_{YY}^{YY,XX}(e_{12}=\ & -\beta_3\beta_1^{-1}e_{21}e_{XX},\\
\d_{YY}^{YY,XX}(e_{21})=\ & -\beta_3\beta_1^{-1}e_{12}e_{XX},\\
\d_{YY}^{YY,XX}(e_{22})=\ & \beta_3\beta_1^{-1}e_{11}e_{XX},\\
\d_{YY}^{YY,YY}(e_{11})=\ & \beta_5\beta_4^{-1}\lp e_{11}e_{22}+e_{12}e_{21}+e_{21}e_{12}+e_{22}e_{11}\rp,\\
\d_{YY}^{YY,YY}(e_{12})=\ & \beta_5\beta_4^{-1}\lp e_{11}e_{21}-e_{12}e_{22}+e_{21}e_{11}-e_{22}e_{12}\rp,\\
\d_{YY}^{YY,YY}(e_{21})=\ & \beta_5\beta_4^{-1}\lp e_{11}e_{12}+e_{12}e_{11}-e_{21}e_{22}-e_{22}e_{21}\rp,\\
\d_{YY}^{YY,YY}(e_{22})=\ & \beta_5\beta_4^{-1}\lp e_{11}e_{11}-e_{12}e_{12}-e_{21}e_{21}+e_{22}e_{22}\rp.
\end{align}

Now we would like to gauge $11+XX+YY$.  We can view this as again associated to the regular representation, with identifications
\begin{align}
e=\ & v_1+v_b+v_{b^2}+v_a+v_{ab}+v_{ab^2},\\
e_{XX}=\ & \frac{c_{XX}}{\sqrt{6}}\lp v_1+v_b+v_{b^2}-v_a-v_{ab}-v_{ab^2}\rp,\\
e_{11}=\ & \frac{c_{YY}}{2\sqrt{3}}\lp -2v_1+v_b+v_{b^2}+2v_a-v_{ab}-v_{ab^2}\rp,\\
e_{12}=\ & \frac{c_{YY}}{2}\lp v_b-v_{b^2}-v_{ab}+v_{ab^2}\rp,\\
e_{21}=\ & \frac{c_{YY}}{2}\lp -v_b+v_{b^2}-v_{ab}+v_{ab^2}\rp,\\
e_{22}=\ & \frac{c_{YY}}{2\sqrt{3}}\lp -2v_1+v_b+v_{b^2}-2v_a+v_{ab}+v_{ab^2}\rp.
\end{align}
These are the same vectors as in the $\Rep(S_3)$ case with identifications $e_{XX}=e_X$, $e_{11}=e_{Y_21}$, $e_{12}=e_{Y_11}$, $e_{21}=e_{Y_22}$, $e_{22}=e_{Y_12}$ and $c_{XX}=c_X$, $c_{YY}=c_1=c_2$.  In particular, the actions of $\m$ and $\Delta$ are exactly as before, and so this leads us to 
\be
\m_{11,XX}^{XX}=\m_{11,YY}^{YY}=\m_{XX,11}^{XX}=\m_{YY,11}^{YY}=1,\quad\m_{XX,XX}^{11}=\frac{c_{XX}^2}{6\beta_1},\quad\m_{XX,YY}^{YY}=\frac{c_{XX}}{\sqrt{6}\beta_2},\non
\ee
\be
\m_{YY,XX}^{YY}=\frac{c_{XX}}{\sqrt{6}\beta_3},\quad\m_{YY,YY}^{11}=\frac{c_{YY}^2}{6\beta_4},\quad\m_{YY,YY}^{XX}=\frac{c_{YY}^2}{\sqrt{6}c_{XX}\beta_6},\qquad\m_{YY,YY}^{YY}=-\frac{c_{YY}}{2\sqrt{3}\beta_5},
\ee
and
\be
\Delta_{11}^{11,11}=\Delta_{XX}^{11,XX}=\Delta_{XX}^{XX,11}=\Delta_{YY}^{11,YY}=\Delta_{YY}^{YY,11}=\frac{1}{6},\quad\Delta_{11}^{XX,XX}=\frac{\beta_1}{c_{XX}^2},\non
\ee
\be
\Delta_{11}^{YY,YY}=\frac{\beta_4}{c_{YY}^2},\quad\Delta_{XX}^{YY,YY}=\frac{c_{XX}\beta_4}{\sqrt{6}c_{YY}\beta_2},\quad\Delta_{YY}^{XX,YY}=\frac{\beta_4}{\sqrt{6}c_{XX}\beta_6},\non
\ee
\be
\Delta_{YY}^{YY,XX}=\frac{\beta_1}{\sqrt{6}c_{XX}\beta_3},\quad\Delta_{YY}^{YY,YY}=-\frac{\beta_4}{2\sqrt{3}c_{YY}\beta_5}.
\ee
Putting it together gives the partition function
\begin{align}
\label{eq:DiagPF}
Z_{11+XX+YY}=\ & \frac{1}{6}\ls Z_{11,11}^{11}+Z_{11,XX}^{XX}+Z_{11,YY}^{YY}+Z_{XX,11}^{XX}+Z_{XX,XX}^{11}+\frac{\beta_1}{\beta_2\beta_3}Z_{XX,YY}^{YY}\right.\non\\
& \qquad\left. +Z_{YY,11}^{YY}+\frac{\beta_4}{\beta_3\beta_5}Z_{YY,XX}^{YY}+Z_{YY,YY}^{11}+\frac{\beta_4}{\beta_2\beta_6}Z_{YY,YY}^{XX}+\frac{\beta_4}{2\beta_5^2}Z_{YY,YY}^{YY}\rs.
\end{align}

\subsubsection{Consistency Check}
\label{sec:cosistency}

We can quickly check that this result is consistent with (\ref{fc_diag}).  Suppose we obtained our $\Rep(S_3\times S_3)$ theory by gauging $S_3\times S_3$ for a product theory $A\otimes B$, with each factor having an $S_3$ symmetry.  In this case the partial traces above all factorize, e.g.~
\be
Z_{XX,YY}^{YY}=Z_{X,Y}^{(A/S_3)\,Y}Z_{X,Y}^{(B/S_3)\,Y}.
\ee
Now for the $A/S_3$ and $B/S_3$ partial traces, we can write them in terms of $A$ and $B$ partial traces by using a combination of knowing how the untwisted sector works, modular invariance, and the knowledge that the $1+Y$ orbifold of say $A/S_3$ should give $A/\Z_2$.  The result is
\begin{align}
\label{pt_duality1}
Z_{1,1}^1=\ & \frac{1}{6}\lp Z_{1,1}+2Z_{1,b}+3Z_{1,a}+2Z_{b,1}+2Z_{b,b}+2Z_{b,b^2}+3Z_{a,1}+3Z_{a,a}\rp,\\
Z_{1,X}^X=\ & \frac{1}{6}\lp Z_{1,1}+2Z_{1,b}+3Z_{1,a}+2Z_{b,1}+2Z_{b,b}+2Z_{b,b^2}-3Z_{a,1}-3Z_{a,a}\rp,\\
Z_{1,Y}^Y=\ & \frac{1}{6}\lp 2Z_{1,1}+4Z_{1,b}+6Z_{1,a}-2Z_{b,1}-2Z_{b,b}-2Z_{b,b^2}\rp,\\
Z_{X,1}^X=\ & \frac{1}{6}\lp Z_{1,1}+2Z_{1,b}-3Z_{1,a}+2Z_{b,1}+2Z_{b,b}+2Z_{b,b^2}+3Z_{a,1}-3Z_{a,a}\rp,\\
Z_{X,X}^1=\ & \frac{1}{6}\lp Z_{1,1}+2Z_{1,b}-3Z_{1,a}+2Z_{b,1}+2Z_{b,b}+2Z_{b,b^2}-3Z_{a,1}+3Z_{a,a}\rp,\\
Z_{Y,1}^Y=\ & \frac{1}{6}\lp 2Z_{1,1}-2Z_{1,b}+4Z_{b,1}-2Z_{b,b}-2Z_{b,b^2}+6Z_{a,1}\rp,\\
Z_{Y,Y}^1=\ & \frac{1}{6}\lp 2Z_{1,1}-2Z_{1,b}-2Z_{b,1}-2Z_{b,b}+4Z_{b,b^2}+6Z_{a,a}\rp,\\
Z_{Y,Y}^Y=\ & \frac{2\beta_5^2}{3\beta_4}\lp Z_{1,1}-Z_{1,b}-Z_{b,1}+2Z_{b,b}-Z_{b,b^2}\rp,\\
Z_{X,Y}^Y=\ & \frac{\beta_2\beta_3}{3\beta_1}\lp -Z_{1,1}-2Z_{1,b}+3Z_{1,a}+Z_{b,1}+Z_{b,b}+Z_{b,b^2}\rp,\\
Z_{Y,X}^Y=\ & \frac{\beta_3\beta_6}{3\beta_4}\lp -Z_{1,1}+Z_{1,b}-2Z_{b,1}+Z_{b,b}+Z_{b,b^2}+3Z_{a,1}\rp,\\
Z_{Y,Y}^X=\ & \frac{\beta_2\beta_6}{3\beta_4}\lp Z_{1,1}-Z_{1,b}-Z_{b,1}-Z_{b,b}+2Z_{b,b^2}-3Z_{a,a}\rp.
\label{pt_duality2}
\end{align}
As a check on these, we have
\begin{align}
Z_{1+X}=\ & \hlf\ls Z_{1,1}^1+Z_{1,X}^X+Z_{X,1}^X+Z_{X,X}^1\rs\non\\
=\ & \frac{1}{3}\ls Z_{1,1}+2Z_{1,b}+2Z_{b,1}+2Z_{b,b}+2Z_{b,b^2}\rs=Z_{\Z_3},\\
Z_{1+Y}=\ & \frac{1}{3}\ls Z_{1,1}^1+Z_{1,Y}^Y+Z_{Y,1}^Y+Z_{Y,Y}^1+\frac{\beta_4}{2\beta_5^2}Z_{Y,Y}^Y\rs\non\\
=\ & \hlf\ls Z_{1,1}+Z_{1,a}+Z_{a,1}+Z_{a,a}\rs=Z_{\Z_2},\\
Z_{1+X+2Y}=\ & \frac{1}{6}\ls Z_{1,1}^1+Z_{1,X}^X+2Z_{1,Y}^Y+Z_{X,1}^X+Z_{X,X}^1-\frac{2\beta_1}{\beta_2\beta_3}Z_{X,Y}^Y\right.\non\\
& \qquad\left. +2Z_{Y,1}^Y-\frac{2\beta_4}{\beta_3\beta_6}Z_{Y,X}^Y+2Z_{Y,Y}^1+\frac{2\beta_4}{\beta_2\beta_6}Z_{Y,Y}^X+\frac{2\beta_4}{\beta_5^2}Z_{Y,Y}^Y\rs\non\\
=\ & Z_{1,1}.
\end{align}
Now we apply these decompositions to the $\Rep(S_3)\boxtimes\Rep(S_3)$ orbifold partition function (\ref{eq:DiagPF}) and we find precisely the $S_3$ orbifold partition function (\ref{eq:S3DiagPF}), in perfect agreement with (\ref{fc_diag}).

\section{Application to TQFT}
\label{sec:tqft}

Now we should be able to explicitly confirm some of the calculations in \cite[section A.3]{GaugeDefects}.  The setup here is that we would like to gauge the diagonal algebra of the direct product of a theory carrying Rep$(S_3)$ symmetry with a Rep$(S_3)$-symmetric TQFT.  There are four minimal unitary\footnote{Going forward, we will not repeatedly specify that we are working with unitary TQFTs, but it should be understood that we are imposing this restriction. Similarly, when we claim to give an exhaustive list of TQFTs carrying a certain symmetry, we of course mean that it is an exhaustive list of minimal theories (i.e. ones which cannot be written as a direct sum of other minimal TQFTs). Finally, we will talk here about TQFTs rather than passing to gapped phases (equivalence classes of TQFTs differing by Euler counterterms \cite[section 3.2]{bhardwaj2023gapped}) since the duality relations we employ will (in principle, though we will not do computations at higher genus where these terms would show up) determine the Euler terms of a TQFT from its dual.} TQFTs carrying this symmetry \cite{bhardwaj2023gapped}, all of which can be obtained by gauging SPT phases carrying Rep$(S_3)$ or $S_3$ symmetry:
\begin{itemize}
	\item SPT(Rep$(S_3)$), the Rep$(S_3)$ symmetric SPT state.  It has a single ground state in which all of $\Rep(S_3)$ acts trivially.
	\item SPT$(S_3)$/$\Z_2$, a theory with two ground states which we will call the `$X$ SSB' (spontaneous symmetry breaking) phase.  Here $X$ serves to exchange the two ground states and $Y$ acts on either ground state to produce the sum of both.
	\item SPT(Rep$(S_3)$)/$(1+Y)$, the `$Y$ SSB' phase which has three ground states.  $X$ acts trivially on all three ground states and $Y$ acts on any of the three to produce the sum of the remaining two.
	\item SPT$(S_3)$/$S_3$, which is $S_3$ gauge theory.  Like the phase above it has three ground states -- one for each $S_3$ irrep.  The action of Rep$(S_3)$ on the three ground states $\Pi_1$, $\Pi_2$, $\Pi_3$ is \cite{TopOps,bhardwaj2023gapped}
	\be
	\label{rs3_action_on_dw}
	\begin{tabular}{l l l}
		$1\Pi_1 = \Pi_1$ & $1\Pi_2 = \Pi_2$ & $1\Pi_3 = \Pi_3$ \\
		$X\Pi_1 = \Pi_2$ & $X\Pi_2 = \Pi_1$ & $X\Pi_3 = \Pi_3$ \\
		$Y\Pi_1 = \frac{1}{2}\Pi_3$ & $Y\Pi_2 = \frac{1}{2}\Pi_3$ & $Y\Pi_3 = 2\Pi_1 + 2\Pi_2 + \Pi_3$.
	\end{tabular}
	\ee
\end{itemize}
The main approach will be to use the duality between these $\Rep(S_3)$-symmetric TQFTs and the four $S_3$-symmetric ones. Specifically, by gauging the full $S_3$ in the four $S_3$-symmetric TQFTs, we can produce the four $\Rep(S_3)$-symmetric TQFTs.  The $S_3$-symmetric theories in question are:
\begin{itemize}
	\item $\SPT(S_3)$, the $S_3$-symmetric SPT state, with a single ground state.
	\item $\SPT(\Rep(S_3))/(1 + X)$, the ‘$\Z_2$ SSB’ phase with two ground states. The $\Z_2$ exchanges the ground states, while the $\Z_3$ acts trivially in both.
	\item $\SPT(S_3)/\Z_3$, the ‘$\Z_3$ SSB’ phase with three ground states. Here the $\Z_3$ exchanges the three ground states, but the $\Z_2$ does not act totally trivially -- each of the three conjugate $\Z_2$ subgroups of $S_3$ exchanges two of the three ground states, leaving the third fixed.
	\item $\SPT(\Rep(S_3))/(1+X+2Y)$, $\Rep(S_3)$ gauge theory. This phase has six ground states which carry a free action of $S_3$.
\end{itemize}
It is straightforward to see that one produces the former list from the latter by gauging $S_3$, with the only possible complication being the need to identify composed gaugings such as
\be
[\SPT(S_3)/\Z_3]/S_3 = \SPT(S_3)/\Z_2
\ee
and
\be
[\SPT(\Rep(S_3))/(1+X)]/S_3 = \SPT(\Rep(S_3))/(1+Y).
\ee

\subsection{Calculation of $S_3$ Partial Traces}

Using the theta/gauge defect method of \cite{UNI} and \cite{GaugeDefects}, we expect that we should be able to construct the genus one partition functions for all four gaugings of a $\Rep(S_3)$-symmetric theory by taking the product of that theory with each of the four TQFTs described above and gauging the diagonal algebra.  Per the results of section~\ref{sec:diag}, in order to implement this strategy we need to know the partial traces that one would obtain by gauging $\Rep(S_3)$ in each of the four TQFTs.  As mentioned above, it will be easier to first calculate the corresponding partial traces in the four $S_3$-symmetric TQFTs, then using the duality between $S_3$ and $\Rep(S_3)$ to construct the $\Rep(S_3)$ partial traces.\\

Using the notation of section~\ref{sec:s3_gauge}, the four gaugings of $S_3$ are:
\begin{itemize}
	\item Gauging all of $S_3$, with partition function
	\be
	\frac{1}{6}[Z_{1,1}+3(Z_{1,a}+Z_{a,1}+Z_{a,a})+2(Z_{1,b}+Z_{b,1}+Z_{b,b}+Z_{b,b^2})],
	\ee
	corresponding to the $S_3$ SPT phase.
	\item Gauging the $\Z_3$ subgroup of $S_3$, with partition function
	\be
	\frac{1}{3}[Z_{1,1}+2\lp Z_{1,b}+Z_{b,1}+Z_{b,b}+Z_{b,b^2}\rp],
	\ee
	corresponding to the $\Z_2$ SSB phase.
	\item Gauging any of the three conjugate $\Z_2$ subgroups, with partition function
	\be
	\frac{1}{2}[Z_{1,1}+Z_{1,a}+Z_{a,1}+Z_{a,a}],
	\ee
	corresponding to the $\Z_3$ SSB phase.
	\item The trivial gauging, with partition function
	\be
	Z_{1,1},
	\ee
	corresponding to $\Rep(S_3)$ gauge theory.
\end{itemize}

Comparing the four partition functions above with the partition function (\ref{eq:S3DiagPF}) of a diagonal $S_3$ gauging, we can readily identify the partial traces one would obtain from gauging the $S_3$ symmetry of the four TQFTs in question, collected in Table~\ref{table:s3_pt}.
\begin{table}
	\centering
	\begin{tabular}{c | c | c | c | c}
		& SPT($S_3$) & $\Z_2$ SSB & $\Z_3$ SSB & Rep$(S_3)$ Gauge Theory \\\hline
		$Z_{1,1}$ & 1 & 2 & 3 & 6\\
		$Z_{1,a}$ & 1 & 0 & 1 & 0\\
		$Z_{a,1}$ & 1 & 0 & 1 & 0\\
		$Z_{a,a}$ & 1 & 0 & 1 & 0\\
		$Z_{1,b}$ & 1 & 2 & 0 & 0\\
		$Z_{b,1}$ & 1 & 2 & 0 & 0\\
		$Z_{b,b}$ & 1 & 2 & 0 & 0\\
		$Z_{b,b^2}$ & 1 & 2 & 0 & 0\\
	\end{tabular}
	\caption{Partial traces for gaugings of $S_3$-symmetric TQFTs.}
	\label{table:s3_pt}
\end{table}

We can quickly check that this data is consistent with the description given for each of these phases.  The untwisted partition function $Z_{1,1}$, in each case, simply counts the number of ground states in the theory.  The twisted partition functions in each modular orbit then count the number of ground states invariant under that symmetry.\footnote{One way to see this is to recall that $Z_{1,g}$ corresponds to implementing $g$-twisted boundary conditions on a toroidal worldsheet.  When $g$ acts to exchange one ground state with another, any state living in a such a twisted sector would need to extend between disjoint terms in the direct sum decomposition of the theory into its components.  In such a decomposition, the identity operator on the total space is given by the sum of identity operators on each component or `universe' \cite{sharpe2022introduction}, which means that we can calculate any correlation function of local operators in the theory as a whole as the sum of its value in each universe.  Inserting this sum of local identity operators will kill any state extending between universes, so these states cannot contribute.  Therefore the value of the partial trace must be determined purely by the universes in which $g$ acts trivially.}  For instance, the SPT phase has a single ground state invariant under the entire $S_3$ symmetry, and its partial traces are all unity, as expected from a (group-like) SPT.  Conversely, since $S_3$ acts freely on the six ground states of Rep$(S_3)$ gauge theory, no ground states are invariant under any group element insertions.  We also recover the fact that both of the $\Z_2$ SSB phase's ground states are invariant under the order three element $b$ and are exchanged by the order two element $a$.  Similarly, each order two element fixes only one of the three ground states in the $\Z_3$ SSB phase, and as a consequence we have $Z_{1,a}=Z_{a,1}=Z_{a,a}=1$.

\subsection{Calculation of Rep$(S_3)$ Partial Traces}

Now we would like to transform the data of Table~\ref{table:s3_pt} to equivalent data for the four $\Rep(S_3)$-symmetric TQFTs.  Since these are related to their $S_3$ counterparts by gauging, we will use the relations (\ref{pt_duality1})-(\ref{pt_duality2}) to produce the $\Rep(S_3)$ partial traces from the $S_3$ ones.  Doing so results in Table~\ref{table:rs3_pt}.

\begin{table}
	\centering
	\begin{tabular}{c | c | c | c | c}
		& SPT(Rep($S_3$)) & X SSB & Y SSB & $S_3$ Gauge Theory \\\hline
		$Z_{1,1}^1$ & 1 & 2 & 3 & 3\\
		$Z_{1,X}^X$ & 1 & 0 & 3 & 1\\
		$Z_{X,1}^X$ & 1 & 0 & 3 & 1\\
		$Z_{X,X}^1$ & 1 & 0 & 3 & 1\\
		$Z_{1,Y}^Y$ & 2 & 2 & 0 & 1\\
		$Z_{Y,1}^Y$ & 2 & 2 & 0 & 1\\
		$Z_{Y,Y}^1$ & 2 & 2 & 0 & 1\\
		$Z_{X,Y}^Y$ & -2$\frac{\beta_2\beta_3}{\beta_1}$ & 0 & 0 & $\frac{\beta_2\beta_3}{\beta_1}$\\
		$Z_{Y,X}^Y$ & -2$\frac{\beta_3\beta_6}{\beta_4}$ & 0 & 0 & $\frac{\beta_3\beta_6}{\beta_4}$\\
		$Z_{Y,Y}^X$ & 2$\frac{\beta_2\beta_6}{\beta_4}$ & 0 & 0 & $\frac{\beta_2\beta_6}{\beta_4}$\\
		$Z_{Y,Y}^Y$ & 4$\frac{\beta_5^2}{\beta_4}$ & 2$\frac{\beta_5^2}{\beta_4}$ & 0 & 0\\
	\end{tabular}
	\caption{Partial traces for gaugings of Rep$(S_3)$-symmetric minimal unitary TQFTs.}
	\label{table:rs3_pt}
\end{table}
Up to factors of $\beta$ arising from the fact that we have left the $\Rep(S_3)$ associator in a general gauge, the results here are once again integers.  Again we have $Z_{1,1}^1$ counting the number of ground states in each phase.  We can immediately see an oddity of the non-invertible symmetry: the partial traces for the SPT phase do not all take unit value, and in fact some of them are negative, neither of which would happen for group-like symmetries (without cocycle twists).  As a cross-check, the $\SPT(\Rep(S_3))$ partial traces were calculated by a different method in \cite[section 5.4.1]{Notes1}, with identical results.  Additionally, one should keep in mind that the data of Table~\ref{table:rs3_pt} is constrained by modular invariance, which for non-invertible symmetries can relate linear combinations of partial traces.  The full list of $\Rep(S_3)$ modular transformations are presented in \cite[section 3.1.3]{Notes1}; here we reproduce the ones involving linear combinations:
\begin{align}
	Z_{Y,Y}^1(\tau+1)=\ & \hlf Z_{Y,1}^Y(\tau) + \frac{\beta_4}{2 \beta_3 \beta_6} Z_{Y,X}^Y(\tau) + \frac{\beta_4}{2 \beta_5^2} Z_{Y,Y}^Y(\tau),\\
	Z_{Y,Y}^X(\tau+1)=\ & -\frac{\beta_2 \beta_6}{2 \beta_4} Z_{Y,1}^Y(\tau) - \frac{\beta_2}{2 \beta_3} Z_{Y,X}^Y(\tau) + \frac{\beta_2 \beta_6}{2 \beta_5^2} Z_{Y,Y}^Y(\tau),\\
	Z_{Y,Y}^Y(\tau+1)=\ & \frac{\beta_5^2}{\beta_4} Z_{Y,1}^Y(\tau) - \frac{\beta_5^2}{\beta_3 \beta_6} Z_{Y,X}^Y(\tau)\\
	Z_{Y,Y}^1(-1/\tau)=\ & \hlf Z_{Y,Y}^1(\tau) - \frac{\beta_4}{2 \beta_2 \beta_6} Z_{Y,Y}^X(\tau) + \frac{\beta_4}{2 \beta_5^2} Z_{Y,Y}^Y(\tau),\\
	Z_{Y,Y}^X(-1/\tau)=\ & -\frac{\beta_2 \beta_6}{2 \beta_4} Z_{Y,Y}^1(\tau) + \hlf Z_{Y,Y}^X(\tau) + \frac{\beta_2 \beta_6}{2 \beta_5^2} Z_{Y,Y}^Y(\tau),\\
	Z_{Y,Y}^Y(-1/\tau)=\ & \frac{\beta_5^2}{\beta_4} Z_{Y,Y}^1(\tau) + \frac{\beta_5^2}{\beta_2 \beta_6} Z_{Y,Y}^X(\tau).
\end{align}
One can check that the data given in Table~\ref{table:rs3_pt} satisfies these relations.

\subsection{$\Rep(S_3)$ Gaugings from TQFT}

Now we can use the above results for TQFT partition functions in conjunction with the diagonal algebra to produce the genus one partition functions for the gaugings of $\Rep(S_3)$.  For these purposes, let $\mcT$ be a 2d theory carrying Rep$(S_3)$ symmetry.  We would like to take the product of $\mcT$ with the four Rep$(S_3)$-symmetric minimal unitary TQFTs defined at the start of the section and gauge the diagonal algebra in the result.  Denoting this algebra as before by $\mcA$, we combine the data of Table~\ref{table:rs3_pt} with the diagonal algebra partition function (\ref{eq:DiagPF}) to find 
\begin{multline}
	\label{gd1}
	Z([\mcT\otimes\SPT(\Rep(S_3))]/\mcA)=\frac{1}{6}\bigg[ Z_{1,1}^1+Z_{1,X}^X+Z_{X,1}^X+Z_{X,X}^1+2(Z_{1,Y}^Y+Z_{Y,1}^Y+Z_{Y,Y}^1)\\
	+2\left(-\frac{\beta_1}{\beta_2 \beta_3}Z_{X,Y}^Y-\frac{\beta_4}{\beta_3 \beta_6}Z_{Y,X}^Y+\frac{ \beta_4}{\beta_2 \beta_6}Z_{Y,Y}^X+\frac{\beta_4}{\beta_5^2}Z_{Y,Y}^Y\right)\bigg].
\end{multline}
for the SPT phase, 
\be
\label{gd2}
Z([\mcT\otimes\SPT(S_3)/\Z_2]/\mcA) = \frac{1}{3}\ls Z_{1,1}^1+Z_{1,Y}^Y+Z_{Y,1}^Y+Z_{Y,Y}^1+\frac{\beta_4}{2\beta_5^2}Z_{Y,Y}^Y\rs
\ee
for the $X$ SSB phase,
\be
\label{gd3}
Z([\mcT\otimes\SPT(\Rep(S_3))/(1+Y)]/\mcA)=\frac{1}{2}\biggl[Z_{1,1}^1+Z_{1,X}^X+Z_{X,1}^X+Z_{X,X}^1\biggr]
\ee
for the $Y$ SSB phase and
\begin{multline}
\label{gd4}
Z([\mcT\otimes\SPT(S_3)/S_3]/\mcA) = \frac{1}{6}\bigg[ 3Z_{1,1}^1+Z_{1,X}^X+Z_{X,1}^X+Z_{X,X}^1+Z_{1,Y}^Y+Z_{Y,1}^Y+Z_{Y,Y}^1\\
+\frac{\beta_1}{\beta_2\beta_3}Z_{X,Y}^Y+\frac{\beta_4}{\beta_3\beta_6}Z_{Y,X}^Y-\frac{\beta_4}{\beta_2\beta_6}Z_{Y,Y}^X \bigg]
\end{multline}
for $S_3$ gauge theory.\\

These are almost exactly the results we expect.  In (\ref{gd1}), (\ref{gd2}) and (\ref{gd3}) we see the $1+X+2Y$, $1+Y$ and $1+X$ gaugings of $\mcT$, respectively \cite{Notes1}.  The odd one out is (\ref{gd4}), which is not immediately recognizable as one of the Rep$(S_3)$ gaugings (or a linear combination thereof). 

Fortunately, with a little care we can resolve this apparent puzzle.  Let us assume that the right-hand side of (\ref{gd4}) corresponds to a sensible gauging for a Rep$(S_3)$-symmetric theory.  Let us further present that theory as an $S_3$ orbifold.  We can then use the partial trace relations of section~\ref{sec:cosistency} to figure out what happens when we compose the gauging given in (\ref{gd4}) with the $S_3$ orbifold.  Plugging in, the result is in fact $Z_{1,1}^1$ -- the $S_3$ orbifold partition function.  That is, the gauging appearing on the right-hand side of (\ref{gd4}) is in fact the trivial gauging, masquerading as something more complicated.  This phenomenon is known as Morita equivalence -- there can (and generically do) exist distinct Frobenius algebras over fusion categories which, when gauged, give equivalent results \cite[Definition 7.8.17]{EGNO}\cite[section 4.6]{BhardwajTachikawa}.\footnote{Note further that things more or less had to work out this way; in order to get the trivial Rep$(S_3)$ gauging on the nose from the diagonal gauging, we would have needed a Rep$(S_3)$-symmetric TQFT with $Z_{1,1}^1=6$ and all other partial traces vanishing, i.e.~a theory with six ground states admitting a `free' action of Rep$(S_3)$.  But this was meant to be dual under gauging to the $S_3$-symmetric SPT phase, and 2d gauge theory for non-abelian groups does not produce a totally symmetry-broken phase (notably, the number of ground states matches the number of conjugacy classes instead of the order of the group).  There had to be some additional relation to reconcile these facts, and Morita equivalence is what saves the day.}

A second way to show explicitly that the right-hand side of (\ref{gd4}) is equal to $Z_{1,1}^1$ is by starting with the torus partition function and nucleating a small $Y$ loop, then using repeated swap moves with insertions of crossing kernels.  Since this operation should be equivalent to multiplying the partition function by the quantum dimension of the line forming the loop, this procedure results in non-trivial relations between the various partial traces which can be used to simplify (\ref{gd4}).  The details can be found in Appendix~\ref{app:PTRelations}.

There is in fact a third, even more direct way to demonstrate this Morita equivalence.  If $\mathcal{A}$ is a Frobenius algebra in a fusion category $\mathcal{C}$, and $L$ is some other object in the category whose orientation reversal is $\ov{L}$, then one can give a Frobenius algebra structure to $(L\otimes\mathcal{A})\otimes\ov{L}$, and these two algebras will be Morita equivalent.  This construction is briefly reviewed in Appendix~\ref{app:MoritaEquivalentAlgebra}.  We claim that (\ref{gd4}) arises from taking the trivial gauging $\mathcal{A}=1$, and constructing a Morita equivalent (so still physically trivial) gauging by conjugating with $L=1+Y$.  Indeed, let $e$, $e_1$, and $e_2$ be basis vectors for the representation $L=\ov{L}=1+Y$.  Then $\mathcal{A}'=(L\otimes\mathcal{A})\otimes\ov{L}$ is a nine-dimensional representation with basis (in hopefully obvious notation)
\be
\left\{eee,eee_1,eee_2,e_1ee,e_1ee_1,e_1ee_2,e_2ee,e_2ee_1,e_2ee_2\right\}.
\ee
Since the multiplication in $\mathcal{A}$ is trivial ($\m(ee)=e$) and since the evaluation map is simply $\e_1(ee)=1$, $\e_Y(e_ie_j)=\beta_4\d_{ij}$, the multiplication in $\mathcal{A}'$, as explained in (\ref{eq:RepGMoritaMult}), will be given by $\m'(eee,v)=\m(v,eee)=v$ and
\be
\m'(eee_i,eee_j)=0,\quad\m'(eee_i,e_jee)=\beta_4\d_{ij}eee,\quad\m'(eee_i,e_jee_k)=\beta_4\d_{ij}eee_k,\non
\ee
\be
\m'(e_iee,eee_j)=e_iee_j,\quad\m'(e_iee,e_jee)=0,\quad\m'(e_iee,e_jee_k)=0,\non
\ee
\be
\m'(e_iee_j,eee_k)=0,\quad\m'(e_iee_j,e_kee)=\beta_4\d_{jk}e_iee,\quad\m'(e_iee_j,e_kee_\ell)=\beta_4\d_{jk}e_iee_\ell,
\ee
where $i,j,k,\ell$ can run over $1,2$.

The co-multiplication is similarly
\begin{align}
\Delta'(eee)=\ & \frac{1}{3}\ls eee\otimes eee+\beta_4^{-1}\sum_{i=1}^2eee_i\otimes e_iee\rs,\non\\
\Delta'(eee_i)=\ & \frac{1}{3}\ls eee\otimes eee_i+\beta_4^{-1}\sum_{j=1}^2eee_j\otimes e_jee_i\rs,\non\\
\Delta'(e_iee)=\ & \frac{1}{3}\ls e_iee\otimes eee+\beta_4^{-1}\sum_{j=1}^2e_iee_j\otimes e_jee\rs,\non\\
\Delta'(e_iee_j)=\ & \frac{1}{3}\ls e_iee\otimes eee_j+\beta_4^{-1}\sum_{k=1}^2e_iee_k\otimes e_kee_j\rs.
\end{align}

In order to express the $\mathcal{A}'$ orbifold partition function in terms of partial traces, we must break it up into irreps of $S_3$.  We have $\mathcal{A}'=2+X+3Y$, and we can define basis vectors
\begin{align}
E^{(1_1)}=\ & eee,\\
E^{(1_2)}=\ & e_1ee_1+e_2ee_2,\\
E^{(X)}_X=\ & e_1ee_2-e_2ee_1,\\
E^{(Y_1)}_1=\ & eee_1,\\
E^{(Y_1)}_2=\ & eee_2,\\
E^{(Y_2)}_1=\ & e_1ee,\\
E^{(Y_2)}_2=\ & e_2ee,\\
E^{(Y_3)}_1=\ & e_1ee_2+e_2ee_1,\\
E^{(Y_3)}_2=\ & e_1ee_1-e_2ee_2.
\end{align}
In terms of this basis, we can compute the multiplication and compare to our standard fusions (\ref{eq:FirstFusion})-(\ref{eq:LastFusion})
\be
\m_{1_1,1_1}^{\prime\,1_1}=1,\quad\m_{1_1,Y_1}^{\prime\,Y_1}=1,\quad\m_{1_2,1_2}^{\prime\,1_2}=\beta_4,\quad\m_{1_2,X}^{\prime\,X}=\beta_4,\quad\m_{1_2,Y_2}^{\prime\,Y_2}=\beta_4,\quad\m_{1_2,Y_3}^{\prime\,Y_3}=\beta_4,\non
\ee
\be
\m_{X,1_2}^{\prime\,X}=\beta_4,\quad\m_{X,X}^{\prime\,1_2}=-\beta_4\beta_1^{-1},\quad\m_{X,Y_2}^{\prime\,Y_2}=-\beta_4\beta_2^{-1},\quad\m_{X,Y_3}^{\prime\,Y_3}=\beta_4\beta_2^{-1},\non
\ee
\be
\quad\m_{Y_1,1_2}^{\prime\,Y_1}=\beta_4,\quad\m_{Y_1,X}^{\prime\,Y_1}=\beta_4\beta_3^{-1},\quad\m_{Y_1,Y_2}^{\prime\,1_1}=1,\quad\m_{Y_1,Y_3}^{\prime\,Y_1}=\beta_4\beta_5^{-1},\non
\ee
\be
\m_{Y_2,1_1}^{\prime\,Y_2}=1,\quad\m_{Y_2,Y_1}^{\prime\,1_2}=\hlf\beta_4^{-1},\quad\m_{Y_2,Y_1}^{\prime\,X}=\hlf\beta_6^{-1},\quad\m_{Y_2,Y_1}^{\prime\,Y_3}=\hlf\beta_5^{-1},\non
\ee
\be
\m_{Y_3,1_2}^{\prime\,Y_3}=\beta_4,\quad\m_{Y_3,X}^{\prime\,Y_3}=-\beta_4\beta_3^{-1},\quad\m_{Y_3,Y_2}^{\prime\,Y_2}=\beta_4\beta_5^{-1},\quad\m_{Y_3,Y_3}^{\prime\,1_2}=1,\quad\m_{Y_3,Y_3}^{\prime\,X}=-\beta_4\beta_6^{-1},
\ee
with all other components vanishing.  Similarly
\be
\Delta_{1_1}^{\prime\,1_1,1_1}=\frac{1}{3},\quad\Delta_{1_1}^{\prime\,Y_1,Y_2}=\frac{1}{3},\quad\Delta_{1_2}^{\prime\,1_2,1_2}=\frac{1}{6}\beta_4^{-1},\quad\Delta_{1_2}^{\prime\,X,X}=-\frac{1}{6}\beta_4^{-1}\beta_1,\quad\Delta_{1_2}^{\prime\,Y_2,Y_1}=\frac{1}{3}\beta_4,\non
\ee
\be
\Delta_{1_2}^{\prime\,Y_3,Y_3}=\frac{1}{6},\quad\Delta_X^{\prime\,1_2,X}=\frac{1}{6}\beta_4^{-1},\quad\Delta_X^{\prime\,X,1_2}=\frac{1}{6}\beta_4^{-1},\quad\Delta_X^{\prime\,Y_2,Y_1}=-\frac{1}{3}\beta_2^{-1}\beta_4,\non
\ee
\be
\Delta_X^{\prime\,Y_3,Y_3}=\frac{1}{6}\beta_2^{-1},\quad\Delta_{Y_1}^{\prime\,1_1,Y_1}=\frac{1}{3},\quad\Delta_{Y_1}^{\prime\,Y_1,1_2}=\frac{1}{6}\beta_4^{-1},\quad\Delta_{Y_1}^{\prime\,Y_1,X}=-\frac{1}{6}\beta_1\beta_3^{-1}\beta_4^{-1},\non
\ee
\be
\Delta_{Y_1}^{\prime\,Y_1,Y_3}=\frac{1}{6}\beta_5^{-1},\quad\Delta_{Y_2}^{\prime\,1_2,Y_2}=\frac{1}{6}\beta_4^{-1},\quad\Delta_{Y_2}^{\prime\,X,Y_2}=\frac{1}{6}\beta_6^{-1},\quad\Delta_{Y_2}^{\prime\,Y_2,1_1}=\frac{1}{3},\quad\Delta_{Y_2}^{\prime\,Y_3,Y_2}=\frac{1}{6}\beta_5^{-1},\non
\ee
\be
\Delta_{Y_3}^{\prime\,1_2,Y_3}=\frac{1}{6}\beta_4^{-1},\quad\Delta_{Y_3}^{\prime\,X,Y_3}=-\frac{1}{6}\beta_6^{-1},\quad\Delta_{Y_3}^{\prime\,Y_2,Y_1}=\frac{1}{3}\beta_5^{-1}\beta_4,\non
\ee
\be
\Delta_{Y_3}^{\prime\,Y_3,1_2}=\frac{1}{6}\beta_4^{-1},\quad\Delta_{Y_3}^{\prime\,Y_3,X}=\frac{1}{6}\beta_1\beta_3^{-1}\beta_4^{-1}.
\ee

Finally, the partition function is obtained by
\be
Z_{L1\ov{L}}=\sum_{A,B,C}\m_{A,B}^{\prime\,C}\Delta_C^{\prime\,B,A}\,Z_{A,B}^C,
\ee
and this is given by (grouping isomorphic irreps together)
\begin{align}
Z_{L1\ov{L}}=\ & \frac{1}{6}\ls 3Z_{1,1}^1+Z_{1,X}^X+Z_{X,1}^X+Z_{X,X}^1+Z_{1,Y}^Y+Z_{Y,1}^Y+Z_{Y,Y}^1\right.\non\\
& \quad\left. +\frac{\beta_1}{\beta_2\beta_3}Z_{X,Y}^Y+\frac{\beta_4}{\beta_3\beta_6}Z_{Y,X}^Y-\frac{\beta_4}{\beta_2\beta_6}Z_{Y,Y}^X\rs.
\end{align}
This is exactly the combination that appears in (\ref{gd4}).\\

Knowing that the gauging in question is Morita trivial, we can rewrite (\ref{gd4}) as
\be
Z([\mcT\otimes\SPT(S_3)/S_3]/\mcA) = Z_{1,1}^1.
\ee
This completes the desired picture: the four Rep$(S_3)$-symmetric TQFTs are in one-to-one correspondence with the four gaugings of Rep$(S_3)$ symmetry.  Equivalently, we can view the addition of the TQFT as providing an insertion of a theta/gauge defect into the resulting $S_3$-symmetric theory, in which case the SPT phase acts as the identity and the other three theories produce the three non-trivial gaugings of $S_3$.

\section{Discussion}
\label{sec:discussion}

There are multiple natural directions in which we could extend these results.  While in this paper we restricted our attention to fusion categories of the form Rep$(G)\boxtimes\Rep(G)$, one could more generally define diagonal algebras over Rep$(\mathcal{H})\boxtimes\Rep(\mathcal{H})$ for any Hopf algebra $\mathcal{H}$.  A duality relation along the lines of (\ref{fc_diag}) should hold in this case, where neither side necessarily need reduce to a group-like gauging.

Another restriction we made was focusing only on theories in two spacetime dimensions.  Gauging a 0-form symmetry given by a group $G$ in a (2+1)d theory produces a 1-form $\Rep(G)$ symmetry.  Gauging this 1-form symmetry will function much like gauging a (non-invertible) 0-form symmetry in 2d.  In particular we could stack this 3d theory with a 2d $\Rep(G)$-symmetric TQFT and gauge the diagonal algebra to obtain the gauged theory with a topological defect inserted.  Thus, the formalism developed here could be just as well used to study theta defects obtained when condensing anyons in 3d.

Finally, the calculations we have presented form a framework with which one can understand discrete torsion in gaugeable non-invertible symmetries.  Following the example of section~\ref{sec:tqft}, one could begin by listing gaugings of some symmetry $G$, from which one can produce the $G$-symmetric TQFT data (the equivalent of Table~\ref{table:s3_pt}).  If one can deduce relations along the lines of (\ref{pt_duality1})-(\ref{pt_duality2}) between $G$ and $\Rep(G)$ partial traces, one can then produce $\Rep(G)$-symmetric TQFT partial traces (as in our Table~\ref{table:rs3_pt}).  From there, simply knowing the form of the partition function for gauging the diagonal algebra in $\Rep(G)\boxtimes\Rep(G)$ is enough to produce all gaugings of a $\Rep(G)$-symmetric theory, including those with isomorphic algebra objects but different Frobenius algebra structures, which one might regard as a generalization of the phenomenon of discrete torsion to $\Rep(G)$ symmetries.  We plan to pursue this line of thought in upcoming work.

\appendix

\section{Internal Homs and Folding}
\label{app:folding}

There is a technique known as the `internal hom' construction for generating algebras in $\mcC\boxtimes\mcC\op$, given in \cite{EGNO} and reviewed in \cite{Remarks}.  Here we give a simple argument, based on the `folding trick' in CFT, as to why the algebra associated with Hom$(\id,\id)$ should be the diagonal one.

Recalling the correspondence between gaugings, boundary conditions and modules \cite{BhardwajTachikawa}, we would like a method to produce boundary conditions for a theory with symmetry $\mcC\boxtimes\mcC\op$.  The setup used in \cite{EGNO} is to regard $\mcC$ by itself as a module over $\mcC\boxtimes\mcC\op$ -- said another way, we would like to view TDLs in a theory with symmetry $\mcC$ as providing boundary conditions for a theory with symmetry $\mcC\boxtimes\mcC\op$.  One can easily see how this works by taking a 2d CFT with symmetry $\mcC$ and an insertion of the TDL $\mcL$ on its worldsheet:
\begin{center}
\begin{tikzpicture}[scale=2]
	\draw (0,0) -- (2,0);
	\draw (2,0) -- (2,1);
	\draw (0,0) -- (0,1);
	\draw (0,1) -- (2,1);
	\draw[->] (1,0) -- (1,0.5);
	\draw (1,0.5) -- (1,1);
	\node at (1.25,0.5) {$\mcL$};
	\node at (1,1.25) {CFT$(\mcC)$};
\end{tikzpicture}
\end{center}
We then employ the `folding trick' to crease the worldsheet along the insertion of $\mcL$ -- any insertions of lines in $\mcC$ appear on the folded side with reverse orientation, so the resulting theory is one with symmetry $\mcC\boxtimes\mcC\op$, and the $\mcL$ line now serves as a boundary:
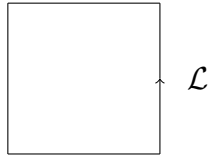
\begin{center}
	\begin{tikzpicture}[scale=2]
		\draw (0,0) -- (1,0);
		\draw (0,0) -- (0,1);
		\draw (0,1) -- (1,1);
		\draw[->] (1,0) -- (1,0.5);
		\draw (1,0.5) -- (1,1);
		\node at (1.25,0.5) {$\mcL$};
		\node at (0.5,1.25) {CFT$(\mcC\boxtimes\mcC\op)$};
	\end{tikzpicture}
\end{center}

The algebra in $\mcC\boxtimes\mcC\op$ generated in this way by $\mcL$ is written as Hom$(\mcL,\mcL)$, and its algebra object will be given by summing over lines which can consistently end on the boundary.  We can find such lines by beginning with all admissible configurations of the form
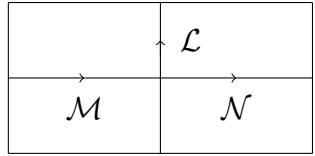
\begin{center}
	\begin{tikzpicture}[scale=2]
		\draw (0,0) -- (2,0);
		\draw (2,0) -- (2,1);
		\draw (0,0) -- (0,1);
		\draw (0,1) -- (2,1);
		\draw[->] (1,0) -- (1,0.75);
		\draw (1,0.65) -- (1,1);
		\draw[->] (0,0.5) -- (0.5,0.5);
		\draw (0.5,0.5) -- (1,0.5);
		\draw[->] (1,0.5) -- (1.5,0.5);
		\draw (1.5,0.5) -- (2,0.5);
		\node at (1.20,0.75) {$\mcL$};
		\node at (1,1.25) {CFT$(\mcC)$};
		\node at (0.5,0.3) {$\mathcal{M}$};
		\node at (1.5,0.3) {$\mathcal{N}$};
	\end{tikzpicture}
\end{center}
for simple lines $\mathcal{M}$ and $\mathcal{N}$ and folding to obtain
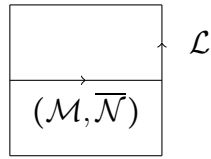
\begin{center}
	\begin{tikzpicture}[scale=2]
		\draw (0,0) -- (1,0);
		\draw (0,0) -- (0,1);
		\draw (0,1) -- (1,1);
		\draw[->] (1,0) -- (1,0.75);
		\draw (1,0.75) -- (1,1);
		\draw[->] (0,0.5) -- (0.5,0.5);
		\draw (0.5,0.5) -- (1,0.5);
		\node at (1.25,0.75) {$\mcL$};
		\node at (0.5,1.25) {CFT$(\mcC\boxtimes\mcC\op)$};
		\node at (0.5,0.3) {$(\mathcal{M},\overline{\mathcal{N}})$};
	\end{tikzpicture}
\end{center}
Let us now set $\mcL=\id$.  Then clearly we must have $\mathcal{N}=\mathcal{M}$, and we obtain a boundary with lines of the form $(\mathcal{M},\overline{\mathcal{M}})$ ending on it:
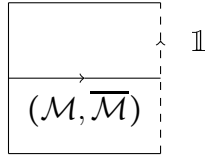
\begin{center}
	\begin{tikzpicture}[scale=2]
		\draw (0,0) -- (1,0);
		\draw (0,0) -- (0,1);
		\draw (0,1) -- (1,1);
		\draw[->,dashed] (1,0) -- (1,0.75);
		\draw[dashed] (1,0.75) -- (1,1);
		\draw[->] (0,0.5) -- (0.5,0.5);
		\draw (0.5,0.5) -- (1,0.5);
		\node at (1.25,0.75) {$\id$};
		\node at (0.5,1.25) {CFT$(\mcC\boxtimes\mcC\op)$};
		\node at (0.5,0.3) {$(\mathcal{M},\overline{\mathcal{M}})$};
	\end{tikzpicture}
\end{center}
This tells us that the algebra object associated with Hom$(\id,\id)$ is
\be
\mcA=\sum_{\mathcal{M}}(\mathcal{M},\overline{\mathcal{M}}),
\ee
exactly as we had found for the diagonal algebra.

\section{Details of $\Rep(S_3)$ Gauging}
\label{app:RepS3Details}

For completeness, we list the explicit bases for fusion and co-fusion intertwiners for $\Rep(S_3)$, 
\begin{align}
\label{eq:FirstFusion}
\la_{X,X}^1(e_Xe_X)=\ & \beta_1e,\\
\la_{X,Y}^Y(e_Xe_1)=\ & \beta_2e_2,\\
\la_{X,Y}^Y(e_Xe_2)=\ & -\beta_2e_1,\\
\la_{Y,X}^Y(e_1e_X)=\ & \beta_3e_2,\\
\la_{Y,X}^Y(e_2e_X)=\ & -\beta_3e_1,\\
\la_{Y,Y}^1(e_1e_1)=\ & \beta_4e,\\
\la_{Y,Y}^1(e_1e_2)=\ & 0,\\
\la_{Y,Y}^1(e_2e_1)=\ & 0,\\
\la_{Y,Y}^1(e_2e_2)=\ & \beta_4e,\\
\la_{Y,Y}^X(e_1e_1)=\ & 0,\\
\la_{Y,Y}^X(e_1e_2)=\ & \beta_6e_X,\\
\la_{Y,Y}^X(e_2e_1)=\ & -\beta_6e_X,\\
\la_{Y,Y}^X(e_2e_2)=\ & 0,\\
\la_{Y,Y}^Y(e_1e_1)=\ & \beta_5e_2,\\
\la_{Y,Y}^Y(e_1e_2)=\ & \beta_5e_1,\\
\la_{Y,Y}^Y(e_2e_1)=\ & \beta_5e_1,\\
\la_{Y,Y}^Y(e_2e_2)=\ & -\beta_5e_2,
\label{eq:LastFusion}
\end{align}
and
\begin{align}
\d_1^{1,1}(e)=\ & ee,\\
\d_1^{X,X}(e)=\ & \beta_1^{-1}e_Xe_X,\\
\d_1^{Y,Y}(e)=\ & \beta_4^{-1}\lp e_1e_1+e_2e_2\rp,\\
\d_X^{1,X}(e_X)=\ & ee_X,\\
\d_X^{X,1}(e_X)=\ & e_Xe,\\
\d_X^{Y,Y}(e_X)=\ & -\beta_2\beta_4^{-1}\lp e_1e_2-e_2e_1\rp,\\
\d_Y^{1,Y}(e_1)=\ & ee_1,\\
\d_Y^{1,Y}(e_2)=\ & ee_2,\\
\d_Y^{X,Y}(e_1)=\ & \beta_6\beta_4^{-1}e_Xe_2,\\
\d_Y^{X,Y}(e_2)=\ & -\beta_6\beta_4^{-1}e_Xe_1,\\
\d_Y^{Y,1}(e_1)=\ & e_1e,\\
\d_Y^{Y,1}(e_2)=\ & e_2e,\\
\d_Y^{Y,X}(e_1)=\ & \beta_3\beta_1^{-1}e_2e_X,\\
\d_Y^{Y,X}(e_2)=\ & -\beta_3\beta_1^{-1}e_1e_X,\\
\d_Y^{Y,Y}(e_1)=\ & \beta_5\beta_4^{-1}\lp e_1e_2+e_2e_1\rp,\\
\d_Y^{Y,Y}(e_2)=\ & \beta_5\beta_4^{-1}\lp e_1e_1-e_2e_2\rp.
\end{align}

Next we need to construct the Frobenius algebra corresponding to the regular representation $1+X+2Y$ and expand its operations in these bases.  We can arrange the vectors in the regular representation by irreps,
\begin{align}
e=\ & v_1+v_b+v_{b^2}+v_a+v_{ab}+v_{ab^2},\\
e_X=\ & \frac{c_X}{\sqrt{6}}\lp v_1+v_b+v_{b^2}-v_a-v_{ab}-v_{ab^2}\rp,\\
e_{Y_11}=\ & \frac{c_1}{2}\lp v_b-v_{b^2}-v_{ab}+v_{ab^2}\rp,\\
e_{Y_12}=\ & \frac{c_1}{2\sqrt{3}}\lp -2v_1+v_b+v_{b^2}-2v_a+v_{ab}+v_{ab^2}\rp,\\
e_{Y_21}=\ & \frac{c_2}{2\sqrt{3}}\lp -2v_1+v_b+v_{b^2}+2v_a-v_{ab}-v_{ab^2}\rp,\\
e_{Y_22}=\ & \frac{c_2}{2}\lp -v_b+v_{b^2}-v_{ab}+v_{ab^2}\rp.
\end{align}
Then the Frobenius multiplication is $\m(v_gv_h)=\d_{g,h}v_g$.  Converting this to the basis above gives $\m(ev)=\m(ve)=v$ along with (and note that $\m(uv)=\m(vu)$),
\begin{align}
\m(e_Xe_X)=\ & \frac{c_X^2}{6}e,\\
\m(e_Xe_{Y_11})=\ & -\frac{c_Xc_1}{\sqrt{6}c_2}e_{Y_22},\\
\m(e_Xe_{Y_12})=\ & \frac{c_Xc_1}{\sqrt{6}c_2}e_{Y_21},\\
\m(e_Xe_{Y_21})=\ & \frac{c_Xc_2}{\sqrt{6}c_1}e_{Y_12},\\
\m(e_Xe_{Y_22})=\ & -\frac{c_Xc_2}{\sqrt{6}c_1}e_{Y_11},\\
\m(e_{Y_11}e_{Y_11})=\ & \frac{c_1^2}{6}e+\frac{c_1}{2\sqrt{3}}e_{Y_12},\\
\m(e_{Y_11}e_{Y_12})=\ & \frac{c_1}{2\sqrt{3}}e_{Y_11},\\
\m(e_{Y_12}e_{Y_12})=\ & \frac{c_1^2}{6}-\frac{c_1}{2\sqrt{3}}e_{Y_12},\\
\m(e_{Y_11}e_{Y_21})=\ & -\frac{c_1}{2\sqrt{3}}e_{Y_22},\\
\m(e_{Y_11}e_{Y_22})=\ & -\frac{c_1c_2}{\sqrt{6}c_X}e_X-\frac{c_1}{2\sqrt{3}}e_{Y_21},\\
\m(e_{Y_12}e_{Y_21})=\ & \frac{c_1c_2}{\sqrt{6}c_X}e_X-\frac{c_1}{2\sqrt{3}}e_{Y_21},\\
\m(e_{Y_12}e_{Y_22})=\ & \frac{c_1}{2\sqrt{3}}e_{Y_22},\\
\m(e_{Y_21}e_{Y_21})=\ & \frac{c_2^2}{6}e-\frac{c_2^2}{2\sqrt{3}c_1}e_{Y_12},\\
\m(e_{Y_21}e_{Y_22})=\ & -\frac{c_2^2}{2\sqrt{3}c_1}e_{Y_11},\\
\m(e_{Y_22}e_{Y_22})=\ & \frac{c_2^2}{6}e+\frac{c_2^2}{2\sqrt{3}c_1}e_{Y_12}.
\end{align}
Now we can expand this in our standard fusion basis and write the multiplication as
\be
\m=\sum_{A,B,C}\m_{A,B}^C\pi_C\circ\la_{R_A,R_B}^{R_C}\circ(\pi_A\otimes\pi_B),
\ee
where $\pi_A$ is simply projection onto the representation $A$ and $R_A$ is the irrep associated to $A$.  This defines coefficients $\m_{A,B}^C$.  Explicitly,
\be
\m_{1,1}^1=\m_{1,X}^X=\m_{1,Y_1}^{Y_1}=\m_{1,Y_2}^{Y_2}=\m_{X,1}^X=\m_{Y_1,1}^{Y_1}=\m_{Y_2,1}^{Y_2}=1,\non
\ee
\be
\m_{X,X}^1=\frac{c_X^2}{6\beta_1},\quad\m_{X,Y_1}^{Y_2}=-\frac{c_Xc_1}{\sqrt{6}c_2\beta_2},\quad\m_{X,Y_2}^{Y_1}=\frac{c_Xc_2}{\sqrt{6}c_1\beta_2},\quad\m_{Y_1,X}^{Y_2}=-\frac{c_Xc_1}{\sqrt{6}c_2\beta_3},\non
\ee
\be
\m_{Y_1,Y_1}^1=\frac{c_1^2}{6\beta_4},\quad\m_{Y_1,Y_1}^{Y_1}=\frac{c_1}{2\sqrt{3}\beta_5},\quad\m_{Y_1,Y_2}^X=-\frac{c_1c_2}{\sqrt{6}c_X\beta_6},\quad\m_{Y_1,Y_2}^{Y_2}=-\frac{c_1}{2\sqrt{3}\beta_5},\non
\ee
\be
\m_{Y_2,X}^{Y_1}=\frac{c_Xc_2}{\sqrt{6}c_1\beta_3},\quad\m_{Y_2,Y_1}^X=\frac{c_1c_2}{\sqrt{6}c_X\beta_6},\quad\m_{Y_2,Y_1}^{Y_2}=-\frac{c_1}{2\sqrt{3}\beta_5},\non
\ee
\be
\m_{Y_2,Y_2}^1=\frac{c_2^2}{6\beta_4},\quad\m_{Y_2,Y_2}^{Y_1}=-\frac{c_2^2}{2\sqrt{3}c_1\beta_5}.
\ee

The co-multiplication is $\Delta(v_g)=v_gv_g$.  Acting on the $e$ vectors, this is
\begin{align}
\Delta(e)=\ & \frac{1}{6}ee+\frac{1}{c_X^2}e_Xe_X+\frac{1}{c_1^2}\lp e_{Y_11}e_{Y_11}+e_{Y_12}e_{Y_12}\rp+\frac{1}{c_2^2}\lp e_{Y_21}e_{Y_21}+e_{Y_22}e_{Y_22}\rp,\\
\Delta(e_X)=\ & \frac{1}{6}\lp ee_X+e_Xe\rp+\frac{c_X}{\sqrt{6}c_1c_2}\lp -e_{Y_11}e_{Y_22}+e_{Y_12}e_{Y_21}+e_{Y_21}e_{Y_12}-e_{Y_22}e_{Y_11}\rp,\\
\Delta(e_{Y_11})=\ & \frac{1}{6}\lp ee_{Y_11}+e_{Y_11}e\rp-\frac{c_1}{\sqrt{6}c_Xc_2}\lp e_Xe_{Y_22}+e_{Y_22}e_X\rp\non\\
& \quad +\frac{1}{2\sqrt{3}c_1}\lp e_{Y_11}e_{Y_12}+e_{Y_12}e_{Y_11}\rp-\frac{c_1}{2\sqrt{3}c_2^2}\lp e_{Y_21}e_{Y_22}+e_{Y_22}e_{Y_21}\rp,\\
\Delta(e_{Y_12})=\ & \frac{1}{6}\lp ee_{Y_12}+e_{Y_12}e\rp+\frac{c_1}{\sqrt{6}c_Xc_2}\lp e_Xe_{Y_21}+e_{Y_21}e_X\rp\non\\
& \quad +\frac{1}{2\sqrt{3}c_1}\lp e_{Y_11}e_{Y_11}-e_{Y_12}e_{Y_12}\rp-\frac{c_1}{2\sqrt{3}c_2^2}\lp e_{Y_21}e_{Y_21}-e_{Y_22}e_{Y_22}\rp,\\
\Delta(e_{Y_21})=\ & \frac{1}{6}\lp ee_{Y_21}+e_{Y_21}e\rp+\frac{c_2}{\sqrt{6}c_Xc_1}\lp e_Xe_{Y_12}+e_{Y_12}e_X\rp\non\\
& \quad -\frac{1}{2\sqrt{3}c_1}\lp e_{Y_11}e_{Y_22}+e_{Y12}e_{Y_21}+e_{Y_21}e_{Y_12}+e_{Y_22}e_{Y_11}\rp,\\
\Delta(e_{Y_22})=\ & \frac{1}{6}\lp ee_{Y_22}+e_{Y_22}e\rp-\frac{c_2}{\sqrt{6}c_Xc_1}\lp e_Xe_{Y_11}+e_{Y_11}e_X\rp\non\\
& \quad -\frac{1}{2\sqrt{3}c_1}\lp e_{Y_11}e_{Y_21}-e_{Y_12}e_{Y_22}+e_{Y_21}e_{Y_11}-e_{Y_22}e_{Y_12}\rp.
\end{align}
Comparing these to the co-fusion basis $\d_{R_1}^{R_2,R_3}$, we can decompose
\be
\Delta=\sum_{A,B,C}\Delta_A^{B,C}(\pi_B\otimes\pi_C)\circ\d_{R_A}^{R_B,R_C}\circ\pi_A,
\ee
finding
\be
\Delta_1^{1,1}=\Delta_X^{1,X}=\Delta_X^{X,1}=\Delta_{Y_1}^{1,Y_1}=\Delta_{Y_1}^{Y_1,1}=\Delta_{Y_2}^{1,Y_2}=\Delta_{Y_2}^{Y_2,1}=\frac{1}{6},\non
\ee
\be
\Delta_1^{X,X}=\frac{\beta_1}{c_X^2},\quad\Delta_1^{Y_1,Y_1}=\frac{\beta_4}{c_1^2},\quad\Delta_1^{Y_2,Y_2}=\frac{\beta_4}{c_2^2},\quad\Delta_X^{Y_1,Y_2}=\frac{c_X\beta_4}{\sqrt{6}c_1c_2\beta_2},\non
\ee
\be
\Delta_X^{Y_2,Y_1}=-\frac{c_X\beta_4}{\sqrt{6}c_1c_2\beta_2},\quad\Delta_{Y_1}^{X,Y_2}=-\frac{c_1\beta_4}{\sqrt{6}c_Xc_2\beta_6},\quad\Delta_{Y_1}^{Y_2,X}=-\frac{c_1\beta_1}{\sqrt{6}c_Xc_2\beta_3},\non
\ee
\be
\Delta_{Y_1}^{Y_1,Y_1}=\frac{\beta_4}{2\sqrt{3}c_1\beta_5},\quad\Delta_{Y_1}^{Y_2,Y_2}=-\frac{c_1\beta_4}{2\sqrt{3}c_2^2\beta_5},\quad\Delta_{Y_2}^{X,Y_1}=\frac{c_2\beta_4}{\sqrt{6}c_Xc_1\beta_6},\non
\ee
\be
\Delta_{Y_2}^{Y_1,X}=\frac{c_2\beta_4}{\sqrt{6}c_Xc_2\beta_6},\quad\Delta_{Y_2}^{Y_1,Y_2}=-\frac{\beta_4}{2\sqrt{3}c_1\beta_5},\quad\Delta_{Y_2}^{Y_2,Y_1}=-\frac{\beta_4}{2\sqrt{3}c_1\beta_5}.
\ee

\section{Bubble Nucleation and Partial Trace Relations}
\label{app:PTRelations}

\begin{figure}
\centering
\begin{tikzpicture}[scale=0.5]
\draw[thick,->] (-2,0) -- (-3.5,2);
\draw[thick] (-3.5,2) -- (-5,4);
\draw[thick,->] (-2,0) -- (-3.5,-2);
\draw[thick] (-3.5,-2) -- (-5,-4);
\draw[thick,->] (2,0) -- (0,0);
\draw[thick] (0,0) -- (-2,0);
\draw[thick,->] (2,0) -- (3.5,2);
\draw[thick] (3.5,2) -- (5,4);
\draw[thick,->] (2,0) -- (3.5,-2);
\draw[thick] (3.5,-2) -- (5,-4);
\draw[thick,color=blue,->,opacity=0.5] (-2.3,0.4) arc (126.87:360:0.5);
\draw[thick,color=blue,->,opacity=0.5] (1.5,0) arc (180:413.13:0.5);
\node at (-5.5,4.5) {$A$};
\node at (-5.5,-4.5) {$B$};
\node at (5.5,-4.5) {$C$};
\node at (5.5,4.5) {$D$};
\node at (0,0.7) {$E$};
\node at (9,0) {$=\ \sum_F\widetilde{K}^{A\,D}_{B\,C}(E,F)\ \times$};
\draw[thick,->] (15,2) -- (13,3.5);
\draw[thick] (13,3.5) -- (11,5);
\draw[thick,->] (15,2) -- (17,3.5);
\draw[thick] (17,3.5) -- (19,5);
\draw[thick,->] (15,2) -- (15,0);
\draw[thick] (15,0) -- (15,-2);
\draw[thick,->] (15,-2) -- (13,-3.5);
\draw[thick] (13,-3.5) -- (11,-5);
\draw[thick,->] (15,-2) -- (17,-3.5);
\draw[thick] (17,-3.5) -- (19,-5);
\draw[thick,color=blue,->,opacity=0.5] (14.6,2.3) arc (143.13:396.87:0.5);
\draw[thick,color=blue,->,opacity=0.5] (14.6,-2.3) arc (216.87:450:0.5);
\node at (10.5,5.5) {$A$};
\node at (10.5,-5.5) {$B$};
\node at (19.5,-5.5) {$C$};
\node at (19.5,5.5) {$D$};
\node at (15.5,0) {$F$};
\end{tikzpicture}
\caption{The conventions for making a swap move.}
\label{fig:SwapMove}
\end{figure}
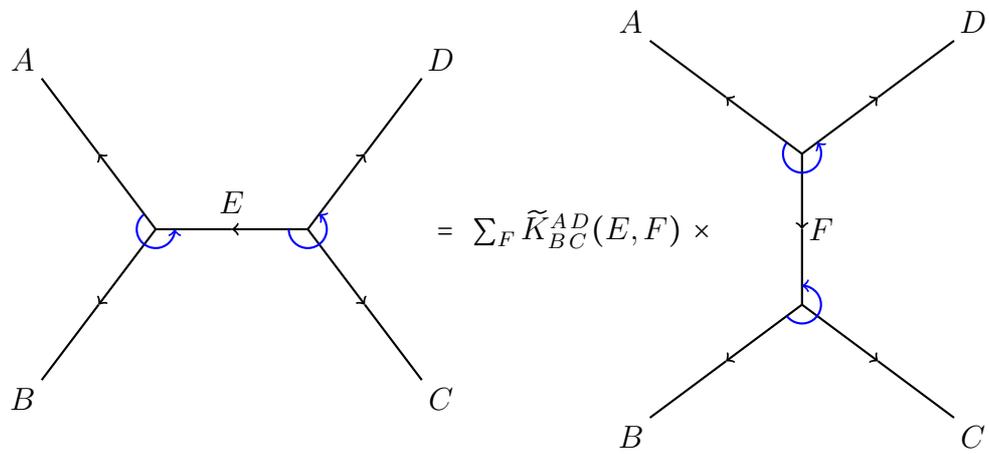

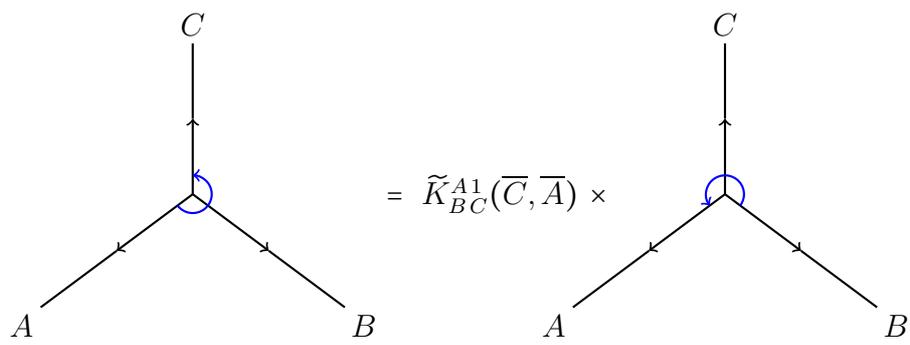
\begin{figure}
\centering
\begin{tikzpicture}[scale=0.5]
\draw[thick,->] (0,0) -- (0,2);
\draw[thick] (0,2) -- (0,4);
\draw[thick,->] (0,0) -- (-2,-1.5);
\draw[thick] (-2,-1.5) -- (-4,-3);
\draw[thick,->] (0,0) -- (2,-1.5);
\draw[thick] (2,-1.5) -- (4,-3);
\draw[thick,color=blue,->,opacity=0.5] (-0.4,-0.3) arc (216.87:450:0.5);
\node at (-4.5,-3.5) {$A$};
\node at (4.5,-3.5) {$B$};
\node at (0,4.5) {$C$};
\node at (8,0) {$=\ \widetilde{K}^{A\,1}_{B\,C}(\ov{C},\ov{A})\ \times$};
\draw[thick,->] (14,0) -- (14,2);
\draw[thick] (14,2) -- (14,4);
\draw[thick,->] (14,0) -- (12,-1.5);
\draw[thick] (12,-1.5) -- (10,-3);
\draw[thick,->] (14,0) -- (16,-1.5);
\draw[thick] (16,-1.5) -- (18,-3);
\draw[thick,color=blue,->,opacity=0.5] (14.4,-0.3) arc (-36.87:216.87:0.5);
\node at (9.5,-3.5) {$A$};
\node at (18.5,-3.5) {$B$};
\node at (14,4.5) {$C$};
\end{tikzpicture}
\caption{As a special case we can permute the ordering at a junction.}
\label{fig:PermuteMove}
\end{figure}

Let's see how the nucleation of a $Y$ line bubble allows us to derive a relation between partial traces in $\Rep(S_3)$.  Before we start, let's review how we can make moves to rearrange networks of topological defect lines within any correlation function.  Our conventions follow those of~\cite{Chang_2019,Notes1},The standard move\footnote{We restrict here to the multiplicity-free case, since this is all we need for $\Rep(S_3)$.} is a swap which introduces factors of the crossing kernel as in Figure~\ref{fig:SwapMove}.  We can also freely attach identity lines wherever we like, with whatever junction ordering we like.  Combining this with our swap move allows us to permute the ordering at a junction, again at the cost of introducing a crossing kernel, as in Figure~\ref{fig:PermuteMove}.

In $\Rep(S_3)$ we have simple lines $1$, $X$, and $Y$, all self-conjugate (i.e.~$\ov{A}=A$).  The crossing kernel components are given by (using the same basis for fusion morphisms as in section~\ref{subsubsec:RepSGauging})~\cite{Notes1}
\be
\widetilde{K}^{1\,C}_{A\,B}(A,C)=\widetilde{K}^{A\,C}_{1\,B}(A,B)=\widetilde{K}^{A\,C}_{B\,1}(C,B)=1,
\ee
and
\begin{align}
\widetilde{K}^{X\,X}_{X\,X}(1,1)=\ & 1,\\
\widetilde{K}^{X\,Y}_{X\,Y}(1,Y)=\ & -\frac{\beta_1}{\beta_2^2},\\
\widetilde{K}^{X\,Y}_{Y\,X}(Y,Y)=\ & 1,\\
\widetilde{K}^{X\,1}_{Y\,Y}(Y,X)=\ & \frac{\beta_2\beta_4}{\beta_1\beta_6},\\
\widetilde{K}^{X\,X}_{Y\,Y}(Y,1)=\ & -\frac{\beta_2\beta_6}{\beta_4},\\
\widetilde{K}^{X\,Y}_{Y\,Y}(Y,Y)=\ & -1,\\
\widetilde{K}^{Y\,Y}_{X\,X}(Y,1)=\ & -\frac{\beta_3^2}{\beta_1},\\
\widetilde{K}^{Y\,1}_{X\,Y}(Y,Y)=\ & -\frac{\beta_3}{\beta_2},\\
\widetilde{K}^{Y\,X}_{X\,Y}(Y,Y)=\ & -\frac{\beta_3}{\beta_2},\\
\widetilde{K}^{Y\,Y}_{X\,Y}(Y,Y)=\ & \frac{\beta_3}{\beta_2},\\
\widetilde{K}^{Y\,1}_{Y\,X}(X,Y)=\ & -\frac{\beta_1\beta_6}{\beta_3\beta_4},\\
\widetilde{K}^{Y\,X}_{Y\,X}(1,Y)=\ & \frac{\beta_4}{\beta_3\beta_6},\\
\widetilde{K}^{Y\,Y}_{Y\,X}(Y,Y)=\ & -1,\\
\widetilde{K}^{Y\,1}_{Y\,Y}(Y,Y)=\ & 1,\\
\widetilde{K}^{Y\,X}_{Y\,Y}(Y,Y)=\ & -1,
\end{align}
\be
\lp\widetilde{K}^{Y\,Y}_{Y\,Y}\rp=\lp\begin{matrix} \hlf & \frac{\beta_4}{2\beta_3\beta_6} & \frac{\beta_4}{2\beta_5^2} \\ -\frac{\beta_2\beta_6}{2\beta_4} & -\frac{\beta_2}{2\beta_3} & \frac{\beta_2\beta_6}{2\beta_5^2} \\ \frac{\beta_5^2}{\beta_4} & -\frac{\beta_5^2}{\beta_3\beta_6} & 0 \end{matrix}\rp,\qquad\lp\widetilde{K}^{Y\,Y}_{Y\,Y}\rp^{-1}=\lp\begin{matrix} \hlf & -\frac{\beta_4}{2\beta_2\beta_6} & \frac{\beta_4}{2\beta_5^2} \\ \frac{\beta_3\beta_6}{2\beta_4} & -\frac{\beta_3}{2\beta_2} & -\frac{\beta_3\beta_6}{2\beta_5^2} \\ \frac{\beta_5^2}{\beta_4} & \frac{\beta_5^2}{\beta_2\beta_6} & 0 \end{matrix}\rp.
\ee

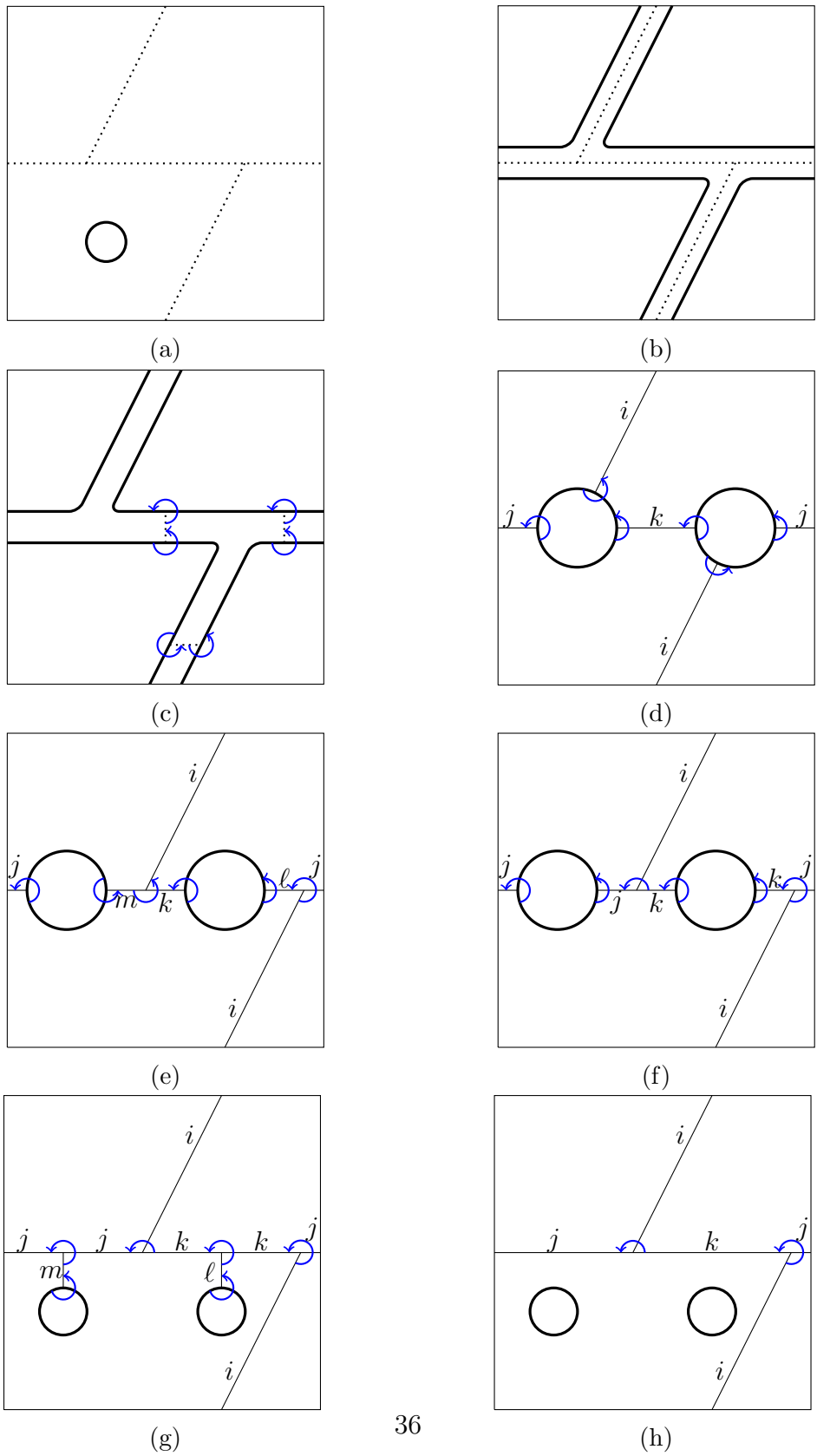
\begin{figure}
\begin{subfigure}{0.5\textwidth}
\centering
\begin{tikzpicture}[scale=0.6]
\draw (0,0) -- (0,8) -- (8,8) -- (8,0) -- (0,0);
\draw[dotted,thick] (0,4) -- (8,4);
\draw[dotted,thick] (4,0) -- (6,4);
\draw[dotted,thick] (2,4) -- (4,8);
\draw[very thick] (2.5,2) circle (0.5);
\end{tikzpicture}
\caption{}
\label{subfig:SmallYBubble}
\end{subfigure}
\begin{subfigure}{0.5\textwidth}
\centering
\begin{tikzpicture}[scale=0.6]
\draw (0,0) -- (0,8) -- (8,8) -- (8,0) -- (0,0);
\draw[dotted,thick] (0,4) -- (8,4);
\draw[dotted,thick] (4,0) -- (6,4);
\draw[dotted,thick] (2,4) -- (4,8);
\draw[very thick,rounded corners] (3.6,0) -- (5.4,3.6) -- (0,3.6);
\draw[very thick,rounded corners] (4.4,0) -- (6.2,3.6) -- (8,3.6);
\draw[very thick,rounded corners] (0,4.4) -- (1.8,4.4) -- (3.6,8);
\draw[very thick,rounded corners] (4.4,8) -- (2.6,4.4) -- (8,4.4);
\end{tikzpicture}
\caption{}
\label{subfig:LargeYBubble}
\end{subfigure}
\begin{subfigure}{0.5\textwidth}
\centering
\begin{tikzpicture}[scale=0.6]
\draw (0,0) -- (0,8) -- (8,8) -- (8,0) -- (0,0);
\draw[very thick,rounded corners] (3.6,0) -- (5.4,3.6) -- (0,3.6);
\draw[very thick,rounded corners] (4.4,0) -- (6.2,3.6) -- (8,3.6);
\draw[very thick,rounded corners] (0,4.4) -- (1.8,4.4) -- (3.6,8);
\draw[very thick,rounded corners] (4.4,8) -- (2.6,4.4) -- (8,4.4);
\draw[dotted,thick] (4.1,1) -- (4.9,1);
\draw[thick,color=blue,->,opacity=0.5] (4.23,1.27) arc (63.43:360:0.3);
\draw[thick,color=blue,->,opacity=0.5] (4.6,1) arc (180:423.43:0.3);
\draw[dotted,thick] (7,3.6) -- (7,4.4);
\draw[thick,color=blue,->,opacity=0.5] (6.7,3.6) arc (180:450:0.3);
\draw[thick,color=blue,->,opacity=0.5] (7,4.1) arc (270:540:0.3);
\draw[dotted,thick] (4,3.6) -- (4,4.4);
\draw[thick,color=blue,->,opacity=0.5] (3.7,3.6) arc (180:450:0.3);
\draw[thick,color=blue,->,opacity=0.5] (4,4.1) arc (270:540:0.3);
\end{tikzpicture}
\caption{}
\label{subfig:ReconnectingLines}
\end{subfigure}
\begin{subfigure}{0.5\textwidth}
\centering
\begin{tikzpicture}[scale=0.6]
\draw (0,0) -- (0,8) -- (8,8) -- (8,0) -- (0,0);
\draw[very thick] (2,4) circle (1);
\draw[very thick] (6,4) circle (1);
\draw (4,0) -- (5.55,3.11);
\draw (2.45,4.89) -- (4,8);
\draw (0,4) -- (1,4);
\draw (3,4) -- (5,4);
\draw (7,4) -- (8,4);
\draw[thick,color=blue,->,opacity=0.5] (1.05,3.70) arc (278.63:540:0.3);
\draw[thick,color=blue,->,opacity=0.5] (2.95,3.70) arc (261.37:458.63:0.3);
\draw[thick,color=blue,->,opacity=0.5] (2.16,4.99) arc (179.32:423.43:0.3);
\draw[thick,color=blue,->,opacity=0.5] (5.05,3.70) arc (278.63:540:0.3);
\draw[thick,color=blue,->,opacity=0.5] (6.95,3.70) arc (261.37:458.63:0.3);
\draw[thick,color=blue,->,opacity=0.5] (5.31,3.28) arc (144.80:342.06:0.3);
\node at (4.2,1) {$i$};
\node at (7.7,4.3) {$j$};
\node at (4,4.3) {$k$};
\node at (0.3,4.3) {$j$};
\node at (3.2,7) {$i$};
\end{tikzpicture}
\caption{}
\label{subfig:PartialWithJunctionBubbles}
\end{subfigure}
\begin{subfigure}{0.5\textwidth}
\centering
\begin{tikzpicture}[scale=0.6]
\draw (0,0) -- (0,8) -- (8,8) -- (8,0) -- (0,0);
\draw[very thick] (1.5,4) circle (1);
\draw[very thick] (5.5,4) circle (1);
\draw (0,4) -- (0.5,4);
\draw (2.5,4) -- (4.5,4);
\draw (6.5,4) -- (8,4);
\draw (3.5,4) -- (5.5,8);
\draw (5.5,0) -- (7.5,4);
\draw[thick,color=blue,->,opacity=0.5] (7.37,3.74) arc (243.43:540:0.3);
\draw[thick,color=blue,->,opacity=0.5] (6.45,3.70) arc (261.37:458.63:0.3);
\draw[thick,color=blue,->,opacity=0.5] (4.55,3.70) arc (278.63:540:0.3);
\draw[thick,color=blue,->,opacity=0.5] (3.2,4) arc (180:423.43:0.3);
\draw[thick,color=blue,->,opacity=0.5] (2.45,4.30) arc (98.63:360:0.3);
\draw[thick,color=blue,->,opacity=0.5] (0.55,3.70) arc (278.63:540:0.3);
\node at (5.7,1) {$i$};
\node at (4.7,7) {$i$};
\node at (7.8,4.6) {$j$};
\node at (0.2,4.6) {$j$};
\node at (7,4.3) {$\ell$};
\node at (4,3.7) {$k$};
\node at (3,3.7) {$m$};
\end{tikzpicture}
\caption{}
\label{subfig:PartialWithEdgeBubbles}
\end{subfigure}
\begin{subfigure}{0.5\textwidth}
\centering
\begin{tikzpicture}[scale=0.6]
\draw (0,0) -- (0,8) -- (8,8) -- (8,0) -- (0,0);
\draw[very thick] (1.5,4) circle (1);
\draw[very thick] (5.5,4) circle (1);
\draw (0,4) -- (0.5,4);
\draw (2.5,4) -- (4.5,4);
\draw (6.5,4) -- (8,4);
\draw (3.5,4) -- (5.5,8);
\draw (5.5,0) -- (7.5,4);
\draw[thick,color=blue,->,opacity=0.5] (7.37,3.74) arc (243.43:540:0.3);
\draw[thick,color=blue,->,opacity=0.5] (6.45,3.70) arc (261.37:458.63:0.3);
\draw[thick,color=blue,->,opacity=0.5] (4.55,3.70) arc (278.63:540:0.3);
\draw[thick,color=blue,->,opacity=0.5] (3.8,4) arc (0:180:0.3);
\draw[thick,color=blue,->,opacity=0.5] (2.45,3.70) arc (261.37:458.63:0.3);
\draw[thick,color=blue,->,opacity=0.5] (0.55,3.70) arc (278.63:540:0.3);
\node at (5.7,1) {$i$};
\node at (4.7,7) {$i$};
\node at (7.8,4.6) {$j$};
\node at (0.2,4.6) {$j$};
\node at (7,4.3) {$k$};
\node at (4,3.7) {$k$};
\node at (3,3.7) {$j$};
\end{tikzpicture}
\caption{}
\label{subfig:PartialWithEdgeBubblesRotated}
\end{subfigure}
\begin{subfigure}{0.5\textwidth}
\centering
\begin{tikzpicture}[scale=0.6]
\draw (0,0) -- (0,8) -- (8,8) -- (8,0) -- (0,0);
\draw[very thick] (1.5,2.5) circle (0.6);
\draw[very thick] (5.5,2.5) circle (0.6);
\draw (0,4) -- (8,4);
\draw (3.5,4) -- (5.5,8);
\draw (5.5,0) -- (7.5,4);
\draw (1.5,4) -- (1.5,3.1);
\draw (5.5,4) -- (5.5,3.1);
\draw[thick,color=blue,->,opacity=0.5] (7.37,3.74) arc (243.43:540:0.3);
\draw[thick,color=blue,->,opacity=0.5] (5.5,3.7) arc (270:540:0.3);
\draw[thick,color=blue,->,opacity=0.5] (3.8,4) arc (0:180:0.3);
\draw[thick,color=blue,->,opacity=0.5] (1.5,3.7) arc (270:540:0.3);
\draw[thick,color=blue,->,opacity=0.5] (5.21,3.03) arc (194.48:450:0.3);
\draw[thick,color=blue,->,opacity=0.5] (1.21,3.03) arc (194.48:450:0.3);
\node at (5.7,1) {$i$};
\node at (4.7,7) {$i$};
\node at (7.8,4.6) {$j$};
\node at (6.5,4.3) {$k$};
\node at (4.5,4.3) {$k$};
\node at (2.5,4.3) {$j$};
\node at (0.5,4.3) {$j$};
\node at (5.2,3.5) {$\ell$};
\node at (1.2,3.5) {$m$};
\end{tikzpicture}
\caption{}
\label{subfig:PartialWithTadpoles}
\end{subfigure}
\begin{subfigure}{0.5\textwidth}
\centering
\begin{tikzpicture}[scale=0.6]
\draw (0,0) -- (0,8) -- (8,8) -- (8,0) -- (0,0);
\draw[very thick] (1.5,2.5) circle (0.6);
\draw[very thick] (5.5,2.5) circle (0.6);
\draw (0,4) -- (8,4);
\draw (3.5,4) -- (5.5,8);
\draw (5.5,0) -- (7.5,4);
\draw[thick,color=blue,->,opacity=0.5] (7.37,3.74) arc (243.43:540:0.3);
\draw[thick,color=blue,->,opacity=0.5] (3.8,4) arc (0:180:0.3);
\node at (5.7,1) {$i$};
\node at (4.7,7) {$i$};
\node at (7.8,4.6) {$j$};
\node at (5.5,4.3) {$k$};
\node at (1.5,4.3) {$j$};
\end{tikzpicture}
\caption{}
\label{subfig:PartialWithFreeBubbles}
\end{subfigure}
\caption{Steps involved in relating two partial trace expressions.}
\end{figure}

We start with the torus partition function, which is equivalent to the partial trace $Z_{1,1}^1$, and add a bubble of $Y$ line somewhere, as in Figure~\ref{subfig:SmallYBubble} (we use dotted lines to represent the identity, thick lines are for $Y$, and ordinary solid lines will have labels which may be summed over).  This bubble simply introduces a factor of the quantum dimension of $Y$, which is $2$.  Next we expand the $Y$ loop out to the locations of the identity line insertions (Figure~\ref{subfig:LargeYBubble}).  We can then redistribute the trivial identity lines, erasing the original ones and putting new ones joining the edges of the $Y$ loop across each of the three straight segments (Figure~\ref{subfig:ReconnectingLines}).  Next, on each of these three segments we perform a swap move.  This results in (Figure~\ref{subfig:PartialWithJunctionBubbles}) a sum over $i$, $j$, and $k$, each running over the simple lines $1$, $X$, and $Y$, and weighted by factors $\widetilde{K}^{Y\,Y}_{Y\,Y}(1,i)\widetilde{K}^{Y\,Y}_{Y\,Y}(1,j)\widetilde{K}^{Y\,Y}_{Y\,Y}(1,k)$.  Note that even though the sum runs over all possible combinations, and there are allowable topological operators we can put at every junction, if a direct $ijk$ junction is not allowed (i.e.~if the identity doesn't appear in that triple fusion product), then the corresponding diagram must in fact vanish.  We can see this by an OPE type argument.  If we take a disk just large enough to contain one of the $Y$ loops, and we use local conformal transformations to shrink the disk to a point, we should get an equivalent diagram with a topological $ijk$ junction; if such a junction doesn't exist (other than the zero operator), then the original diagram must vanish.  So for instance we won't get a contribution with $i=j=k=X$.  With another pair of swap moves we can move the bubbles off the junctions onto lines (Figure~\ref{subfig:PartialWithEdgeBubbles}).  This introduces a sum over $\ell$ and $m$ weighted by $\widetilde{K}^{Y\,Y}_{i\,j}(Y,\ell)(\widetilde{K}^{Y\,i}_{Y\,k})^{-1}(Y,m)$.  However another OPE argument in which we shrink the loops tells us that the diagram will vanish unless $\ell=k$ and $m=j$.  To get to Figure~\ref{subfig:PartialWithEdgeBubblesRotated} we use this fact and also apply two permuation operations, introducing a factor of $\widetilde{K}^{j\,1}_{k\,i}(i,j)\widetilde{K}^{Y\,1}_{Y\,j}(j,Y)$.  Next, another pair of swaps moves the $Y$ bubbles off of the $j$ and $k$ segments (Figure~\ref{subfig:PartialWithTadpoles}).  This produces a sum over $\ell$ and $m$ weighted by $(\widetilde{K}^{Y\,j}_{Y\,j})^{-1}(Y,m)(\widetilde{K}^{Y\,k}_{Y\,k})^{-1}(Y,\ell)$.  Finally, one more loop-shrinking OPE argument tells us that the diagram vanishes unless $\ell=m=1$, in which case we can erase the $\ell$ and $m$ lines leaving two free $Y$ loops that each contribute a factor of $2$, and a simple factor of $Z_{i,j}^k$ (Figure~\ref{subfig:PartialWithFreeBubbles}).

All together, we are left with
\begin{align}
2Z_{1,1}^1=\ & \sum_{i,j,k}\ls\widetilde{K}^{Y\,Y}_{Y\,Y}(1,i)\widetilde{K}^{Y\,Y}_{Y\,Y}(1,j)\widetilde{K}^{Y\,Y}_{Y\,Y}(1,k)\rs\ls\widetilde{K}^{Y\,Y}_{i\,j}(Y,k)\lp\widetilde{K}^{Y\,i}_{Y\,k}\rp^{-1}(Y,j)\rs\non\\
& \qquad\times\ls\widetilde{K}^{j\,1}_{k\,i}(i,j)\widetilde{K}^{Y\,1}_{Y\,j}(j,Y)\rs\ls\lp\widetilde{K}^{Y\,Y}_{Y\,j}\rp^{-1}(Y,1)\lp\widetilde{K}^{Y\,k}_{Y\,k}\rp^{-1}(Y,1)\rs\,2^2\,Z_{i,j}^k\non\\
=\ & \hlf\ls\vphantom{\frac{\beta_4}{\beta_2\beta_6}} Z_{1,1}^1+Z_{1,X}^X+Z_{1,Y}^Y+Z_{X,1}^X+Z_{X,X}^1+Z_{Y,1}^Y+Z_{Y,Y}^1\right.\non\\
& \qquad\left. +\frac{\beta_1}{\beta_2\beta_3}Z_{X,Y}^Y+\frac{\beta_4}{\beta_3\beta_6}Z_{Y,X}^Y-\frac{\beta_4}{\beta_2\beta_6}Z_{Y,Y}^X\rs.
\end{align}
Note that the coefficient of $Z_{Y,Y}^Y$ comes out as zero because $\widetilde{K}^{Y\,Y}_{Y\,Y}(Y,Y)=0$.

Adding $Z_{1,1}^1$ to each side and then dividing by $3$ shows that the expression on the right-hand side of (\ref{gd4}) is precisely equal to $Z_{1,1}^1$, as claimed.  It's clear that similar manipulations can be used to obtain additional relations between partial traces.

\section{Morita Equivalence by Line Conjugation}
\label{app:MoritaEquivalentAlgebra}

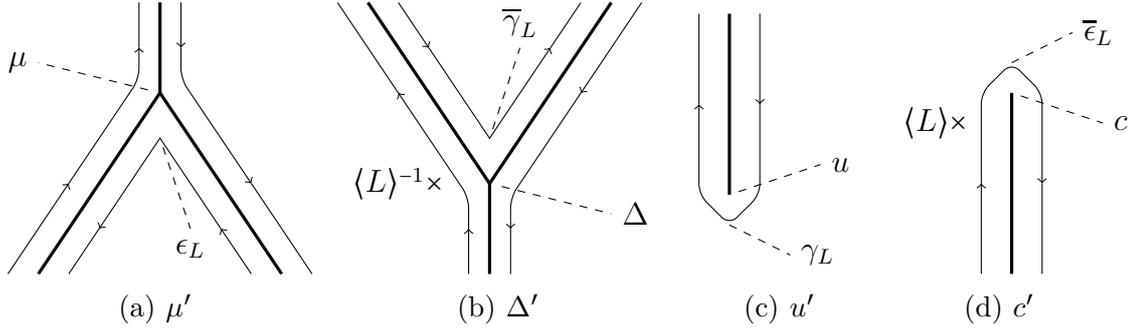
\begin{figure}
\begin{subfigure}{0.3\textwidth}
\centering
\begin{tikzpicture}[scale=0.4]
\draw[very thick] (-4,-6) -- (0,0) -- (0,3);
\draw[very thick] (4,-6) -- (0,0);
\draw (-3,-6) -- (-2,-4.5);
\draw[->] (2,-4.5) -- (0,-1.5) -- (-2,-4.5);
\draw[->] (3,-6) -- (2,-4.5);
\draw[->] (-5,-6) -- (-3,-3);
\draw[->,rounded corners] (-3,-3) -- (-0.69,0.46) -- (-0.69,1.5);
\draw (-0.69,1.5) -- (-0.69,3);
\draw[->] (0.69,3) -- (0.69,1.5);
\draw[->,rounded corners] (0.69,1.5) -- (0.69,0.46) -- (3,-3);
\draw (3,-3) -- (5,-6);
\draw[dashed] (-0.4,0.1) -- (-4,1) node[left] {$\m$};
\draw[dashed] (0.1,-1.8) -- (1,-4.5) node[below] {$\e_L$};
\end{tikzpicture}
\caption{$\m'$}
\label{subfig:muPrime}
\end{subfigure}
\begin{subfigure}{0.3\textwidth}
\centering
\begin{tikzpicture}[scale=0.4]
\draw[very thick] (0,-3) -- (0,0) -- (-4,6);
\draw[very thick] (0,0) -- (4,6);
\draw[->] (-0.69,-3) -- (-0.69,-1.5);
\draw[->,rounded corners] (-0.69,-1.5) -- (-0.69,-0.46) -- (-3,3);
\draw (-3,3) -- (-5,6);
\draw[->] (5,6) -- (3,3);
\draw[->,rounded corners] (3,3) -- (0.69,-0.46) -- (0.69,-1.5);
\draw (0.69,-1.5) -- (0.69,-3);
\draw[->] (-3,6) -- (-2,4.5);
\draw[->] (-2,4.5) -- (0,1.5) -- (2,4.5);
\draw (2,4.5) -- (3,6);
\draw[dashed] (0.4,-0.1) -- (4,-1) node[right] {$\Delta$};
\draw[dashed] (0.1,1.8) -- (1,4.5) node[above] {$\ov{\g}_L$};
\node at (-3,0) {$\langle L\rangle^{-1}\times$};
\end{tikzpicture}
\caption{$\Delta'$}
\label{subfig:DeltaPrime}
\end{subfigure}
\begin{subfigure}{0.19\textwidth}
\centering
\begin{tikzpicture}[scale=0.4]
\draw[very thick] (0,0) -- (0,6);
\draw[->] (1,6) -- (1,3);
\draw[->,rounded corners] (1,3) -- (1,0) -- (0,-1) -- (-1,0) -- (-1,3);
\draw (-1,3) -- (-1,6);
\draw[dashed] (0.3,0.1) -- (3,1) node[right] {$u$};
\draw[dashed] (0,-1) -- (2,-2) node[right] {$\g_L$};
\end{tikzpicture}
\caption{$u'$}
\label{subfig:uPrime}
\end{subfigure}
\begin{subfigure}{0.19\textwidth}
\centering
\begin{tikzpicture}[scale=0.4]
\draw[very thick] (0,-6) -- (0,0);
\draw[->] (-1,-6) -- (-1,-3);
\draw[->,rounded corners] (-1,-3) -- (-1,0) -- (0,1) -- (1,0) -- (1,-3);
\draw (1,-3) -- (1,-6);
\draw[dashed] (0.3,-0.1) -- (3,-1) node[right] {$c$};
\draw[dashed] (0,1) -- (2,2) node[right] {$\ov{\e}_L$};
\node at (-2.5,-1) {$\langle L\rangle\times$};
\end{tikzpicture}
\caption{$c'$}
\label{subfig:cPrime}
\end{subfigure}
\caption{Morita equivalent Frobenius algebra construction.  Thick lines represent $\mathcal{A}$ and thin directed lines represent $L$.  For conventions of evaluation and co-evaluation maps, see~\cite{Notes1}.}
\end{figure}

Suppose we have a fusion category $\mathcal{C}$ in which we have selected a Frobenius algebra.  That is we have an algebra object $\mathcal{A}$ with associated multiplication $\m\in\Hom(\mathcal{A}\otimes\mathcal{A},\mathcal{A})$, co-multiplication $\Delta\in\Hom(\mathcal{A},\mathcal{A}\otimes\mathcal{A})$, unit $u\in\Hom(1,\mathcal{A})$, and co-unit $c\in\Hom(\mathcal{A},1)$, satisfying various compatibility conditions.  Now given another object $L$ in $\mathcal{C}$, we can build a Morita equivalent algebra object $\mathcal{A}'=(L\otimes\mathcal{A})\otimes\ov{L}$.  The multiplication $\m'\in\Hom(\mathcal{A}'\otimes\mathcal{A}',\mathcal{A}')$ is built by combining the multiplication  and co-unit of $\mathcal{A}$ with the associators $\al_{A,B,C}\in\Hom((A\otimes B)\otimes C,A\otimes(B\otimes C))$ and evaluation maps $\e_A\in\Hom(\ov{A}\otimes A,1)$ of $\mathcal{C}$.  Explicitly, we construct $\m'$ by the series of compositions
\begin{multline}
\m':\ \mathcal{A}'\otimes\mathcal{A}'=((L\otimes\mathcal{A})\otimes\ov{L})\otimes((L\otimes\mathcal{A})\otimes\ov{L})\stackrel{\al_{L\mathcal{A},\ov{L},(L\mathcal{A})\ov{L}}}{\longrightarrow}(L\otimes\mathcal{A})\otimes(\ov{L}\otimes((L\otimes\mathcal{A})\otimes\ov{L}))\\
\stackrel{\operatorname{id}_{L\mathcal{A}}\otimes\al^{-1}_{\ov{L},L\mathcal{A},\ov{L}}}{\longrightarrow}(L\otimes\mathcal{A})\otimes((\ov{L}\otimes(L\otimes\mathcal{A}))\otimes\ov{L})\stackrel{\operatorname{id}_{L\mathcal{A}}\otimes(\al^{-1}_{\ov{L},L,\mathcal{A}}\otimes\operatorname{id}_{\ov{L}})}{\longrightarrow}(L\otimes\mathcal{A})\otimes(((\ov{L}\otimes L)\otimes\mathcal{A})\otimes\ov{L})\\
\stackrel{\operatorname{id}_{L\mathcal{A}}\otimes((\e_L\otimes\operatorname{id}_{\mathcal{A}})\otimes\operatorname{id}_{\ov{L}})}{\longrightarrow}(L\otimes\mathcal{A})\otimes((1\otimes\mathcal{A})\otimes\ov{L})\stackrel{\operatorname{id}_{L\mathcal{A}}\otimes((\m\circ(u\otimes\operatorname{id}_{\mathcal{A}}))\otimes\operatorname{id}_{\ov{L}}}{\longrightarrow}(L\otimes\mathcal{A})\otimes(\mathcal{A}\otimes\ov{L})\\
\stackrel{\al^{-1}_{L\mathcal{A},\mathcal{A},\ov{L}}}{\longrightarrow}((L\otimes\mathcal{A})\otimes\mathcal{A})\otimes\ov{L}\stackrel{\al_{L,\mathcal{A},\mathcal{A}}\otimes\operatorname{id}_{\ov{L}}}{\longrightarrow}(L\otimes(\mathcal{A}\otimes\mathcal{A}))\otimes\ov{L}\stackrel{(\operatorname{id}_L\otimes\m)\otimes\operatorname{id}_{\ov{L}}}{\longrightarrow}(L\otimes\mathcal{A})\otimes\ov{L}=\mathcal{A}'.
\end{multline}
Diagramatically, we can represent this (with the many, many associator maps implicit) as in Figure~\ref{subfig:muPrime}.  Here we use a thick line to represent $\mathcal{A}$ and a thin line to represent $L$.  In similar fashion we can construct the co-multiplication, unit, and co-unit for $\mathcal{A}'$ (Figures~\ref{subfig:DeltaPrime}, \ref{subfig:uPrime}, and~\ref{subfig:cPrime} respectively).  One subtlety is that the condition $\m'\circ\Delta'=\operatorname{id}_{\mathcal{A}'}$ requires that $\Delta'$ contain an explicit factor of $\langle L\rangle^{-1}$, where $\langle L\rangle$ is the expectation value of an $L$ loop on the plane.  Then this also requires that the co-unit $c'$ must be given an explicit factor of $\langle L\rangle$.

Because of the large numbers of associator and inverse associator maps, this might seem daunting to implement.  Fortunately, however, in a representation category like $\Rep(G)$, the associator is canonical.  Indeed if $v_A$, $v_B$, and $v_C$ are vectors in the $G$-representations $A$, $B$, and $C$ respectively, then the intertwiner corresponding to the associator homomorphism $\al_{A,B,C}$ simply acts as
\be
\al_{A,B,C}((v_A\otimes v_B)\otimes v_C)=v_A\otimes(v_B\otimes v_C).
\ee
Because of this it is easy to characterize the action of $\m'$.  If $v_i$ are a basis for the representation $L$, $\ov{v}_j$ for $\ov{L}$, and $w_a$ for $\mathcal{A}$, then we can write an arbitrary vector in $(L\otimes\mathcal{A})\otimes\ov{L}$ as $v_iw_a\ov{v}_j$, and we have
\be
\label{eq:RepGMoritaMult}
\m'(v_iw_a\ov{v}_j,v_kw_b\ov{v}_\ell)=\e_L(\ov{v}_j,v_k)\,v_i\m(w_a,w_b)\ov{v}_\ell.
\ee
For co-multiplication, suppose we can write the co-evaluation map $\ov{\g}_L$ and co-multiplication $\Delta$ as
\be
\ov{\g}_L(1)=\sum_{i,j}\ov{g}_{i,j}\ov{v}_jv_i,\qquad\Delta(w_a)=\sum_{b,c}d_a^{b,c}w_bw_c.
\ee
Then
\be
\Delta'(v_iw_au_j)=\frac{1}{\langle L\rangle}\sum_{k,\ell}\sum_{b,c}\ov{g}_{k,\ell}d_a^{b,c}\lp v_iw_b\ov{v}_\ell\rp\otimes\lp v_kw_c\ov{v}_j\rp.
\ee
Similarly,
\be
u'(1)=\sum_{i,j}g_{i,j}\,v_iu(1)\ov{v}_j,\qquad c'(v_iw_a\ov{v}_j)=\langle L\rangle\,\ov{\e}_L(v_i\ov{v}_j)\,c(w_a),
\ee
where the coefficients $g_{i,j}$ come from
\be
\g_L(1)=\sum_{i,j}g_{i,j}\,v_i\ov{v}_j.
\ee

We would also like to show that $\mathcal{A}$ and $\mathcal{A}'$ are in fact Morita equivalent, i.e.~that their categories of left-modules are equivalent.  Indeed, let $M$ be a left-module for $\mathcal{A}$, so $M$ is an object in $\mathcal{C}$ and we have an action $m\in\Hom(\mathcal{A}\otimes M,M)$ that satisfies various compatibility conditions with the Frobenius algebra structure.  Then we can check that $M'=L\otimes M$ is a left-module for $\mathcal{A}'$ with action
\begin{multline}
m':\ \mathcal{A}'\otimes M'=((L\otimes\mathcal{A})\otimes\ov{L})\otimes(L\otimes M)\stackrel{\al_{L\mathcal{A},\ov{L},LM}}{\longrightarrow}(L\otimes\mathcal{A})\otimes(\ov{L}\otimes(L\otimes M))\\
\stackrel{\operatorname{id}_{L\mathcal{A}}\otimes\al^{-1}_{\ov{L},L,M}}{\longrightarrow}(L\otimes\mathcal{A})\otimes((\ov{L}\otimes L)\otimes M)\stackrel{\operatorname{id}_{L\mathcal{A}}\otimes(\e_L\otimes\operatorname{id}_M)}{\longrightarrow}(L\otimes\mathcal{A})\otimes(1\otimes M)\\
\stackrel{\operatorname{id}_{L\mathcal{A}}\otimes(m\circ(u\otimes\operatorname{id}_M))}{\longrightarrow}(L\otimes\mathcal{A})\otimes M\stackrel{\al_{L,\mathcal{A},M}}{\longrightarrow}L\otimes (\mathcal{A}\otimes M)\stackrel{\operatorname{id}_L\otimes m}{\longrightarrow}L\otimes M=M'
\end{multline}
It's not hard to check that this satisfies the necessary conditions.  Similarly, given a left-module $M'$ of $\mathcal{A}'$ with action $m'$, we claim that $M''=\ov{L}\otimes M'$ is a left-module of $\mathcal{A}$ with action
\begin{multline}
m'':\ \mathcal{A}\otimes M''=\mathcal{A}\otimes(\ov{L}\otimes M')\stackrel{((c\otimes\operatorname{id}_{\mathcal{A}})\circ\Delta)\otimes\operatorname{id}_{\ov{L}M'}}{\longrightarrow}(1\otimes\mathcal{A})\otimes(\ov{L}\otimes M')\\
\stackrel{(\ov{\g}_L\otimes\operatorname{id}_{\mathcal{A}})\otimes\operatorname{id}_{\ov{L}M'}}{\longrightarrow}((\ov{L}\otimes L)\otimes\mathcal{A})\otimes(\ov{L}\otimes M')\stackrel{\al_{\ov{L},L,\mathcal{A}}\otimes\operatorname{id}_{\ov{L}M'}}{\longrightarrow}(\ov{L}\otimes(L\otimes\mathcal{A}))\otimes(\ov{L}\otimes M')\\
\stackrel{\al^{-1}_{\ov{L},L\mathcal{A},\ov{L}M'}}{\longrightarrow}\ov{L}\otimes((L\otimes\mathcal{A})\otimes(\ov{L}\otimes M'))\stackrel{\operatorname{id}_{\ov{L}}\otimes\al^{-1}_{L\mathcal{A},\ov{L},M'}}{\longrightarrow}\ov{L}\otimes(((L\otimes\mathcal{A})\otimes\ov{L})\otimes M')\\
=\ov{L}\otimes(\mathcal{A}'\otimes M')\stackrel{\operatorname{id}_{\ov{L}}\otimes m'}{\longrightarrow}\ov{L}\otimes M'=M'',
\end{multline}
again satisfying the required conditions.

\addcontentsline{toc}{section}{References}

\bibliographystyle{utphys}
\bibliography{FCDiag}
	
\end{document}